%% file: main.tex
\documentclass[sigconf]{acmart}
\AtBeginDocument{%
  }

\usepackage{dirtytalk}
\usepackage{multirow}
\usepackage{xcolor}

\usepackage{subcaption}
\usepackage{enumitem}
\usepackage{color}
\usepackage{algorithm}

\definecolor{burgundy}{rgb}{0.5, 0.0, 0.13}
\definecolor{yellow}{rgb}{0.85, 0.65, 0.13}
\definecolor{green}{rgb}{0.0, 0.5, 0.0} 

\definecolor{darkbrown}{rgb}{0.55, 0.27, 0.07}

\newcommand{\revision}[1]{{#1}}
\newcommand{\baseline}[1]{{#1}}
\newcommand{\originality}[1]{{#1}}
\newcommand{\evaluation}[1]{{#1}}
\newcommand{\presentation}[1]{{#1}}
\newcommand{\usecase}[1]{{#1}}

\newcommand{\comments}{1}
\newcommand{\ignore}[1]{}

\ifdefined\comments

\else
\fi

\definecolor{highlight}{RGB}{230, 230, 255} 

\copyrightyear{2025}
\acmYear{2025}
\setcopyright{cc}
\setcctype{by}
\acmConference[CHI '25]{CHI Conference on Human Factors in Computing
Systems}{April 26-May 1, 2025}{Yokohama, Japan}
\acmBooktitle{CHI Conference on Human Factors in Computing Systems (CHI
'25), April 26-May 1, 2025, Yokohama,
Japan}\acmDOI{10.1145/3706598.3713363}
\acmISBN{979-8-4007-1394-1/25/04}

\begin{document}

\title{WhatELSE: Shaping Narrative Spaces at Configurable Level of Abstraction for AI-bridged Interactive Storytelling
}

\author{Zhuoran Lu}
\affiliation{%
  \institution{Department of Computer Science, Purdue University}
  \city{West Lafayette}
  \country{USA}}
\email{lu800@purdue.edu}

\author{Qian Zhou}
\affiliation{%
  \institution{Autodesk Research}
  \city{Toronto}
  \country{Canada}}
\email{qian.zhou@autodesk.com}

\author{Yi Wang}
\affiliation{%
  \institution{Autodesk Research}
  \city{San Francisco}
  \country{USA}}
\email{yi.wang@autodesk.com}

\begin{abstract}

Generative AI significantly enhances player agency in interactive narratives (IN) by enabling just-in-time content generation that adapts to player actions. While delegating generation to AI makes IN more interactive, it becomes challenging for authors to control the space of possible narratives - within which the final story experienced by the player emerges from their interaction with AI. In this paper, we present WhatELSE, an AI-bridged IN authoring system that creates narrative possibility spaces from example stories. WhatELSE provides three views (narrative pivot, outline, and variants) to help authors understand the narrative space and corresponding tools leveraging linguistic abstraction to control the boundaries of the narrative space. Taking innovative LLM-based narrative planning approaches, WhatELSE further unfolds the narrative space into executable game events. Through a user study (N=12) and technical evaluations, we found that WhatELSE enables authors to perceive and edit the narrative space and generates engaging interactive narratives at play-time.


\end{abstract}





\begin{teaserfigure}
\includegraphics[width=0.24 \textwidth]{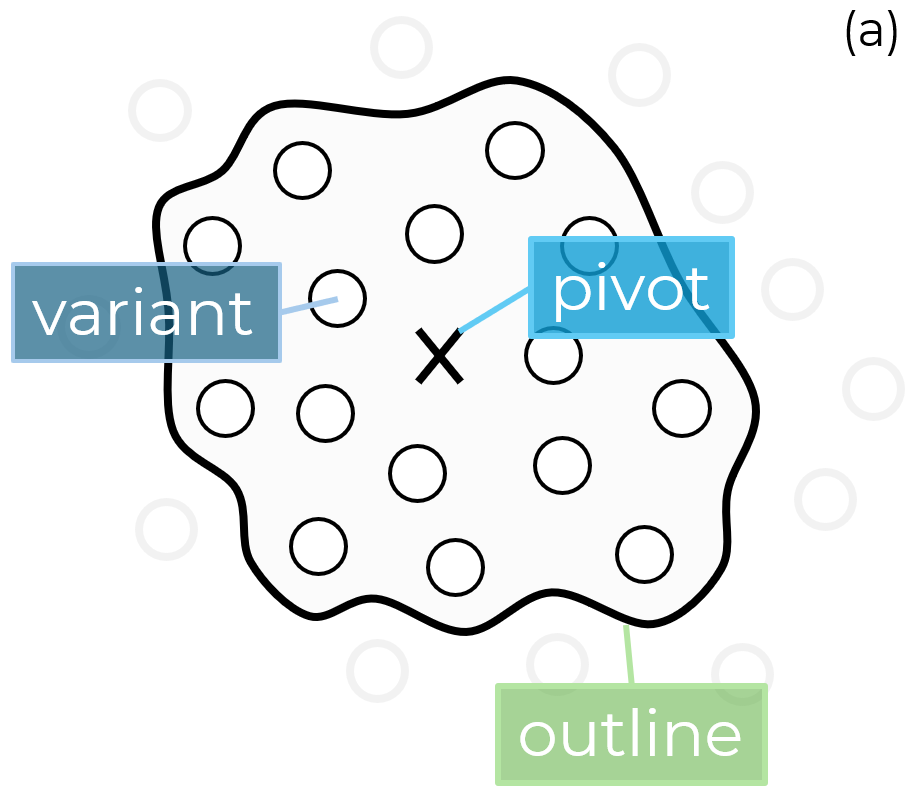}~%
\includegraphics[width=0.24 \textwidth]{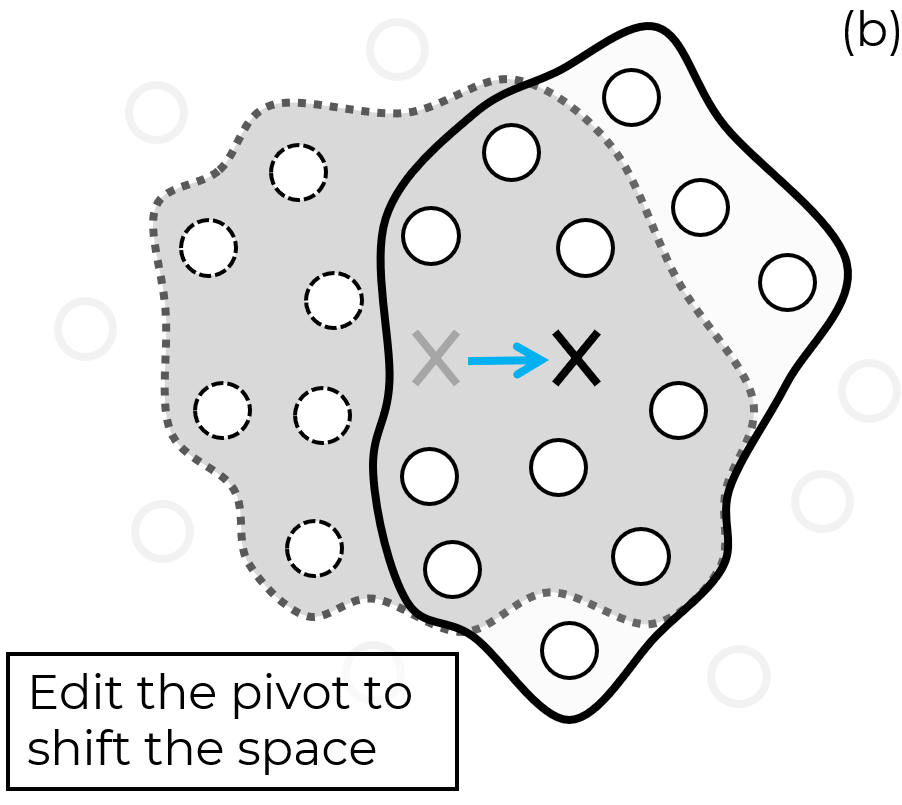}~%
\includegraphics[width=0.24 \textwidth]{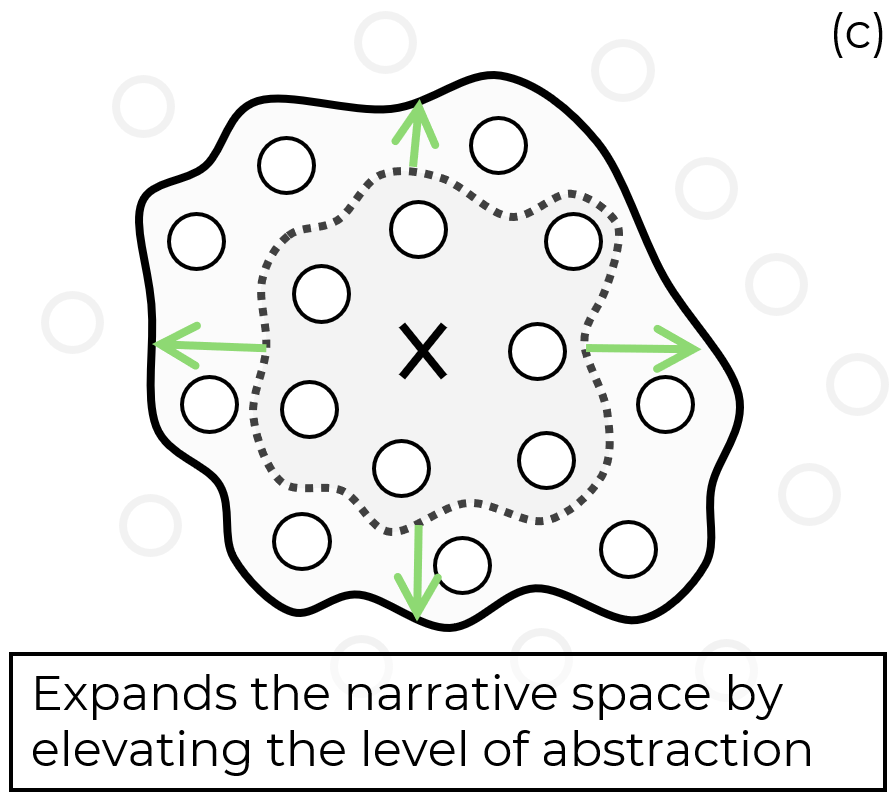}~%
\includegraphics[width=0.24 \textwidth]{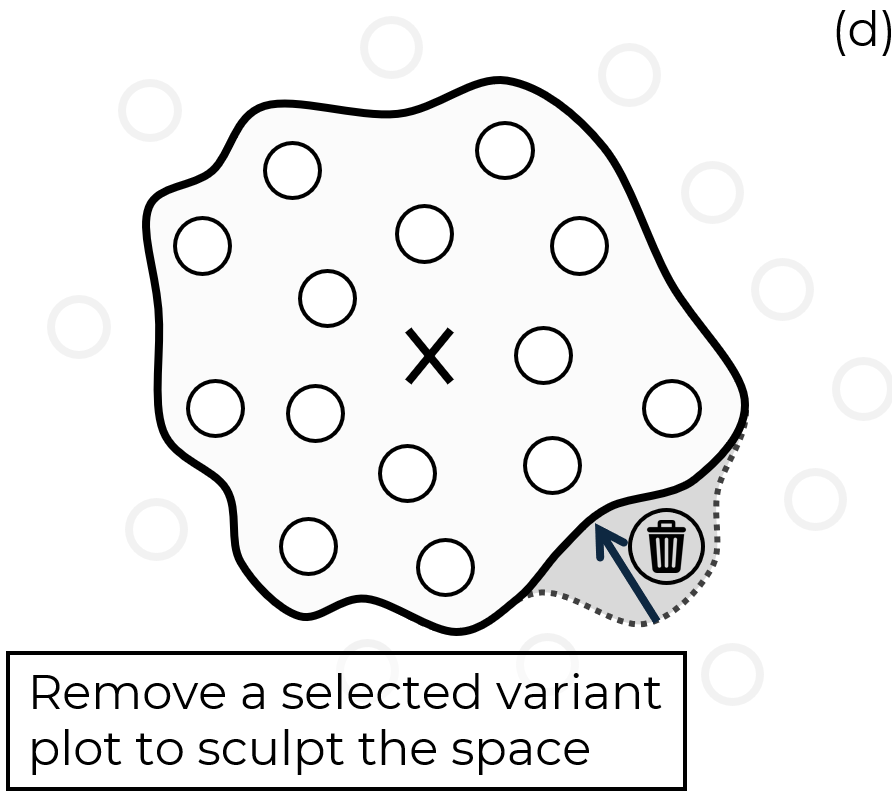}%
\caption{We present WhatELSE, an interactive narrative authoring system that allows users to shape a narrative space using language abstraction. (a) We use the pivot, outline, and variants to describe the narrative space. Users can import an example story as a pivot. The system elevates the pivot into a narrative space. It generates an outline and variants to describe the space. Users can (b) edit the pivot to shift the space, (c) elevate the abstraction level to expand the space, or (d) remove variants to sculpt the space. }
  \label{fig:teaser}
  \Description{}
\end{teaserfigure}


\maketitle
\section{Introduction}
\input{sections/introduction}

\section{Related Work}

\input{sections/relatedWork}




\section{Challenges of AI-Bridged Interactive Narrative Authoring}
\input{sections/concept}


\section{WhatELSE: System Design and Implementation}
\input{sections/system_re}

\section{User Study}
\input{sections/userStudy}

\section{Technical Evaluation}
\input{sections/technicalEvaluation}


\section{Discussion}
\input{sections/discussion}

\begin{acks}
We are grateful to the anonymous reviewers who provided many helpful comments. We also thank our colleagues, David Ledo, Fraser Anderson and Hilmar Koch for feedback and suggestions in preparing the manuscripts and figures. Any opinions, findings, conclusions,
or recommendations expressed here are those of the authors alone.
\end{acks}

\bibliographystyle{ACM-Reference-Format}
\bibliography{llm_narrative}

\end{document}


\title{WhatELSE: Shaping Narrative Spaces at Configurable Level of Abstraction for AI-bridged Interactive Storytelling}
\subtitle{Supplemental Material}

\maketitle
\thispagestyle{empty}

\section{Story Domain and Example Stories Used In User Study}
In the supplementary material, we incorporate a list of files describing the story domain of the Fairytale Forest, and the example stories used in the user study.
\begin{itemize}
    \item \texttt{story\_domain}: the Fairytale Forest story domain, including available characters and their descriptions, available locations, and available actions for characters.
    \item \texttt{example\_narratives}: the narratives we provided to participants as the starting point of their creation in user study. The two example stories each express "kindness is never wasted" and "kindness is not always rewarded", respectively.
\end{itemize}

\section{Prompts Used In Implementation}

We provided the core prompt templates that we used to implement the WhatELSE system and the baseline system in the user study. Specifically, we provide prompts for WhatELSE in the directory named \texttt{whatelse} in the supplementary material:
\begin{itemize}
    \item \texttt{outline\_generation}: the prompt chain to generate outlines with different levels of abstraction, and example pairs of narrative plots with their generated outlines.
    \item \texttt{player\_simulation}: the prompts to simulate different types of player behaviors to drive the narrative progression.
\end{itemize}

Additionally, the prompts used for the baseline system is listed in \texttt{baseline} directory:
\begin{itemize}
    \item \texttt{outline\_generation}: the prompt used for the baseline system to generate an outline.
    \item \texttt{interactivity}: prompting the LLM to serve as a text adventure game engine given a story domain and unfold the game for players to interact under a narrative outline. 
\end{itemize}

\section{Examples of Game Plots}

In \texttt{example\_game\_plot.pdf}, we provide examples of two complete game plots that unfold from the narrative space characterized by a narrative outline consisting of three events. Specifically, the two game plots are driven by different players (Player 1 and Player 2), each taking distinct actions. While what happened in the first event is intentionally set to be the same across both game plots, the narrative progression diverges as each player's actions influence and dynamically adjust the narrative planning thereafter.

\section{Demographic Information of User Study Participants}
In \texttt{user\_study/demographic\_information\_participants.pdf}, we provide the demographic information of participants who attended our user study.

%% file: sections/introduction.tex
Interactive Narrative (IN) is a form of digital storytelling experience where the player can influence a dramatic storyline through their actions \cite{riedl2013interactive,green2014interactive}. 
IN takes various forms in entertainment and education applications, with the most prominent one being branching storylines in role-playing games \cite{riedl2006from}, where the author predefines the range of player actions and creates multiple storylines reflecting the consequences of different player choices. Instead of one single narrative, the author creates a {\em narrative space} consisting of all possible storylines a player can experience. 


The advancement of Large Language Models (LLMs) has the potential to revolutionize IN by enabling the automatic generation of content based on the user's specifications~\cite{li2024pre, li2023synthetic,wei2022chain}. This enables just-in-time generation of narrative content to adapt to different game world states. Instead of enumerating all possible storylines, authors can convey their broad narrative intent to the LLM as prompts, and let the model render concrete narrative instances customized by the player's in-game context \cite{kim2023language,boriskin2024lsg,sweetser2024large,peng2024player}. For example, Inworld Origin \cite{inworldorigin} is a narrative-driven adventure game where the player interacts with LLM-driven characters to solve a mystery. The characters respond to the player with unscripted actions and dialogs that are generated by LLMs at play-time, while still adhering to an overall narrative structure.
AI-bridged IN significantly enhances player agency as the player can now influence the story in ways not restricted by predefined story branches, and also facilitates a form of emergent narrative \cite{aylett1999narrative,suttie2013theoretical} where the final narrative experience can even possibly go beyond the author's anticipation. 

\presentation{To create a traditional IN, the author directly specifies} {\textbf{concrete narrative instances} conditioned on game world states, while in the case of AI-bridged IN, the author has to express their narrative intent as prompts - which requires them to write \textbf{abstract narrative specifications}. \revision{ The abstract specification eventually transforms into concrete narrative instances at play-time based on the player's interaction with the game system.} For example, instead of specifying a story where \textit{``an ant fell into the water and was then saved by a dove by dropping a leaf to the water''}, the author writes in a prompt \textit{``a story where a small creature got into an accident and was saved by another creature''} so that the exact plot can be generated based on the play-time game world states.
Using abstract narrative specification to guide play-time plot progression is an effective way to compactly sketch out a narrative space and impose high-level authorial control.

However, this AI-bridged IN workflow presents challenges both to the author and the AI system, as concrete narrative instances are not accessible at the time of authoring. On one hand, authors struggle to create, perceive, and control the narrative space just by prompting \cite{kreminski2024intent}. Once they write prompts with abstract narrative specification, it is difficult for them to envision specific instances within the defined narrative space \cite{kim2024authors}. On the other hand, it is also difficult for LLMs to unfold the author's narrative intent into a sequence of events that are executable in an external game environment, as LLMs are not trained to simulate the causal dynamics defined by the game mechanism and are known for challenges in maintaining long-term consistency \cite{mirowski2023co}.

To address these challenges, we present {\sc WhatELSE}, an IN authoring system that creates interactive narratives from user-provided example narratives.  
Instead of writing abstract prompts, authors can import narrative instances as {\em pivots} to create a narrative space. The system generates an {\em outline} by abstracting from these instances. To help authors perceive the space, the system uses a simulation process to generate concrete narrative {\em variants} from the outline-defined space. Together, the system uses {\em pivot}, {\em outline}, and {\em variants} to represent and shape the narrative space: (1) authors can shift the space by directly editing the {\em pivot} (Figure~\ref{fig:teaser}.b), (2) they can expand or constrain the narrative space by changing the {\em outline's} level of abstraction (Figure~\ref{fig:teaser}.c), (3) they can also fine-tune the narrative space by removing {\em variants} (Figure~\ref{fig:teaser}.d). 


To support {\sc WhatELSE}, we developed a technical pipeline that supports the bidirectional transformation between outlines and instances using LLM and narrative planning. To generate instances from an outline, we developed a novel LLM-based narrative planning method \revision{following the LLM-Modulo frameworks proposed by Kambhampati et al. \cite{kambhampati2024llms}}, \revision{taking into account possible play-time world states and player behaviors}. Narratives are grounded by character action sequences executable in the game environment and are iteratively generated and reviewed. We leverage an external simulated game environment to guide the validation and revision of plot generation to ensure that the causal dynamics in the game environment are correctly captured by the generated plots. 

This work has three main contributions: 1) an IN authoring system that allows users to shape the narrative space at different levels of abstraction using the outline and instances, 2) a technical pipeline that supports bidirectional transformation between outlines and instances using LLMs and narrative planning; 
and 3) findings from a user study (n=12) and a technical evaluation. Results from the user study showed that \textsc{WhatELSE} helped authors perceive and edit the narrative space. It also demonstrated that the created narrative space could be used to generate engaging, interactive narratives at playtime. Our technical evaluation demonstrated the effectiveness of the pipeline in supporting language abstraction and generating diverse plots that react to player actions.

%% file: sections/relatedWork.tex

\subsection{Authoring Interactive Narratives}

Various tools have been proposed to support authoring Interactive Narratives (IN) in the past decades ~\cite{green2021use,green2018define,roemmele2015creative}. Many of these tools are designed to give better control and management over branching storylines, enabling rich player actions while maintaining authorial control \cite{friedhoff2013untangling}. IN Authoring tools usually organize the narrative space in explicit branching structure  \cite{chen2022does} to enable intuitive understanding, including flowchart-like structures ~\cite{friedhoff2013untangling}, state machines ~\cite{green2020towards,syahputra2019historical}, and collections of modular story chunks conditioned on game world states (``storylet'') \cite{kreminski2018sketching}. 
\color{black}
Every possible narrative instance is manually authored, which ensures that the author’s intent is precisely preserved \cite{riedl2013interactive}. Much of the prior work 
seeks to make the authoring process more efficient through more compact representations of storylines \cite{scigajlo2023generation}. However, despite these efforts, the complexity of defining all possible plot progressions remains a significant challenge \cite{riedl2013interactive}. IN authoring continues to be a time-intensive and engineering-heavy process, often falling short of expectations ~\cite{green2021use,spierling2009authoring}.

Recent efforts have moved towards play-time narrative generation, reducing the need for extensive manual authoring \cite{freiknecht2020procedural, garbe2019storyassembler}. Authors set high-level narrative requirements, allowing for automated procedural story generation that responds to player actions \cite{riedl2006linear}. While this approach saves manual effort, 
authors need to express their authorial intent through means other than high-level specifications. To address this challenge, our system proposes various levels of abstraction grounded in the theory of narrative structure \cite{mckee1997story}, allowing authors to express design requirements from beat-level concrete details to story-level flexible goals.



\color{black}

\subsection{Narrative Generation}

A prominent approach to generating plots is through symbolic narrative planning
\cite{young2013plans, meehan1977tale, lebowitz1985story, riedl2010narrative, ware2011cpocl,ammanabrolu2020story}, which explicitly models the story domain and simulates the causal dynamics of possible plot events to guarantee the causal soundness of generated plots in the simulated context. The author describes a desired world state at the end of plot execution as a ``narrative goal'', and the narrative planning algorithm needs to generate a sequence of state transitions (events) that leads the world state to the narrative goal - called ``narrative plans'' \cite{lebowitz1985story}.


Symbolic narrative planning requires a hand-crafted knowledge base using formal logical language that defines preconditions and effects of a predefined action set within a story \cite{poulakos2016evaluating}. 
Given the extensive engineering work required to construct this knowledge base, the generated plots offer limited complexity and scale.

\originality{The advancement of generative AI also leads to LLM-based methods for plot generation. LLMs can be used to design various narrative elements of the game \cite{lankes2023ai,lanzi2023chatgpt,li2023analyzing}, such as character design~\cite{marincioni2024effect,cox2023conversational,christiansen2024exploring,chiang2024enhancing}, world setting~\cite{jinworldweaver,ratican2024adaptive,short2024designing}, scenes and plots~\cite{chung2022talebrush, yong2023playing, nasir2024word2world,ammanabrolu2020story}. They can also be used at playtime to facilitate just-in-time narrative content generation, such as character dialogs ~\cite{akoury2023framework}. Recent work has also shown that LLMs can be used to drive play-time character behaviors \cite{park2023generative, wang2023humanoid, chen2023agentverse},}
\presentation{leading to plots naturally emerging from the characters.}  \originality{Unlike symbolic narrative planning, which restricts the expression of authorial intent to  formal logical languages, LLM-based methods allow users to write flexible abstract narrative specifications to guide plot generation.}

\originality{
However, LLM-based approaches pose challenges to controllability due to their black-box nature. This lack of control remains a major obstacle for authors looking to adopt LLMs in their workflow \cite{biermann2022tool}. LLM-driven character simulation methods \cite{park2023generative, wang2023humanoid, chen2023agentverse} pose even greater challenges for authorial control due to the emergent behaviors of the LLM-powered characters, leaving a ``herding cat'' problem when generating narratives \cite{ware2021sabre}.
To overcome the challenge, effort has been made to incorporate a symbolic representation of events \cite{ammanabrolu2020story}, introduce sketching input beyond text \cite{chung2022talebrush}, and shift AI to an advisor role \cite{roemmele2015creative}. }

\originality{In this work, we aim to improve the controllability of narrative generation by providing authors with various writing assistance with a configurable level of abstraction in the outline, instance, and sentence level to specify the desired narrative goals and character behaviors. Our system combines symbolic planning, LLMs, and character simulation to generate narratives with causal soundness and emergent behaviors for flexible plot progression.} 

\subsection{Narrative Space}
\originality{The concept of narrative space refers to the range of stories that a system can generate \cite{riedl2006story}, possibly conditioned by constraints or requirements from the author.} A narrative space can be characterized in different ways. For example, traditional interactive fiction defines its narrative space with explicit branching storylines. The symbolic narrative planning system usually defines its narrative space based on the story domain \cite{poulakos2016evaluating}, i.e., the possible events that can happen in a story world derived from the characters, locations, objects, and character action schema. 

\originality{The narrative space of a generative system based on LLMs is characterized by the LLM model itself and the prompt creation mechanism.
Unlike narrative spaces defined by formal representations like branching diagrams and state machines \cite{riedl2013interactive}, narrative spaces based on natural language have soft boundaries due to the inherent fuzziness of natural language semantics. The open-endedness of LLMs makes such narrative spaces more difficult for the author to perceive.} 
\originality{
Existing work has explored semantic abstraction of sentences called ``events'' \cite{ammanabrolu2020story} and ``loglines'' \cite{mirowski2023co}, as a unit of a story that summarizes its central dramatic conflict. Inspired by prior work, we introduce the notion of ``outline'' as the abstract specification that defines the narrative space for generating IN. This ``outline'' has a configurable level of abstraction, allowing the author to adjust the granularity of their authorial control. }

The concept of narrative space is closely related to the notion of design space \cite{biskjaer2014constraint,suh2023structured} or conceptual space \cite{wiggins2006preliminary}. They all refer to a metaphorical space of possibilities - constructed as ideas, designs, solutions, etc. \originality{Previous works have been mostly focusing on exploring a design space in search of specific designs to generate candidates from requirements to establish the space \cite{lin2024design,riedl2006story}, traverse from an existing artifact (pivot) to its alternatives (variants) \cite{matejka2018dream, schulz2017interactive}, and compare these alternatives in multiple dimensions \cite{suh2024luminate,stump2003design,zaman2015gem}. Inspired by prior work, we propose the pivot and variants view to describe the narrative space. }

\originality{Our work differs from prior work by considering the narrative space as the final artifact, rather than an intermediate step in creating a narrative artifact. IN authors create this space for players to explore, making it important for authors to fully understand the narrative space. }
\originality{Creating such space requires balancing between authorial control and emergence \cite{kang2011approach, riedl2013interactive, wang2024storyverse}. On one hand, the space of possible plots needs to be constrained by the author's narrative intent. On the other hand, the space needs to be sufficiently under-constrained so that the player's action and interactions with characters can be reflected in plots that reside in the space. \presentation{ Balancing these two potentially conflicting objectives is the main challenge in AI-bridged IN authoring.} }

%% file: sections/concept.tex

\label{sec:concept}

\label{design_goals}

Interactive Narrative allows players to influence storylines through their actions, with authors creating a narrative space of possible storylines. \revision{During the authoring process, authors predefine the range of player actions and creates multiple storylines reflecting the consequences of different player choices.} AI-bridged IN generates just-in-time narrative content that adapts to different game world states, freeing authors from enumerating storylines. 
However, shifting from traditional IN to AI-bridged IN presents challenges for authors in expressing, perceiving, and controlling the narrative space. Authors often struggle to articulate their implicit narrative intents in high-level prompts \cite{mirowski2023co} and may underexpress their intent to AI systems \cite{kreminski2024intent}. While novice authors might start with a concrete example \cite{tomlinson2006learning,micallef2024use}, a single narrative instance can be both overly detailed and insufficient, as it includes unnecessary specifics and lacks broader context \cite{kreminski2024intent}. Therefore, neither concrete instances nor abstract specifications alone are ideal for defining a narrative space. Instead, the ability to configure the level of abstraction is necessary to support AI-bridged IN authoring. 

On the other hand, once a narrative space is defined via prompts, the author has limited insight into the player experience, as players are responded with unscripted character actions and dialogs generated by LLMs at play-time. It is difficult to identify and prevent the deviations beyond the the author's narrative intent. Therefore, it is important to provide valuable information on how different types of player could react to instances, through which authors could preview the narrative instances as they get unfolded in the player experience \cite{kim2023language}. Furthermore, transforming prompts into meaningful game plots is not trivial. It requires effective narrative planning to generate causally sound event sequences. Central to this success is modeling the logical causal progression of the game plot \cite{riedl2010narrative}. However, LLMs are not natively planners in creating causal progression and have been found to cause hallucinations without external verifier to validate the coherency and executability of the generated plan \cite{kambhampati2024llms}. 
Motivated by the challenges unresolved in AI-bridged IN, we developed the following design goals to guide the design of system:

\textbf{DG1: Enable users to perceive the narrative space.} 
The narrative space in AI-bridged IN contains various possible storylines, which are generated at play-time based on player actions. Authors might struggle to envision what types of variations would be possible. The system should provide representations that can help authors to explore and understand the narrative space. 


\textbf{DG2: Support configurable level of abstraction in editing narrative space.} 
Concrete instances can be overly detailed, while abstract specifications can be too vague. Supporting users to adjust the level of abstraction helps them balance between details and abstraction, which allows the narrative instances to emerge from player interactions, while still adhering to the narrative structure. 


\textbf{DG3: Generate meaningful game events that react to player actions at play-time.} An engaging player experience requires the generated plots to represent logical causal progression that follows the game mechanism.
The proposed system should support simulating casual dynamics and generate meaningful narrative content based on player actions.


%% file: sections/system_re.tex

In this section, we present the interface and features of {\sc WhatELSE}, describe its technical pipeline to facilitate the transformation between narrative instances and narrative outlines, and demonstrate its workflow with an example user story.

\subsection{Narrative Space Editor Interface}

\begin{figure*}[t]
\centering
\includegraphics[width=0.99\textwidth]{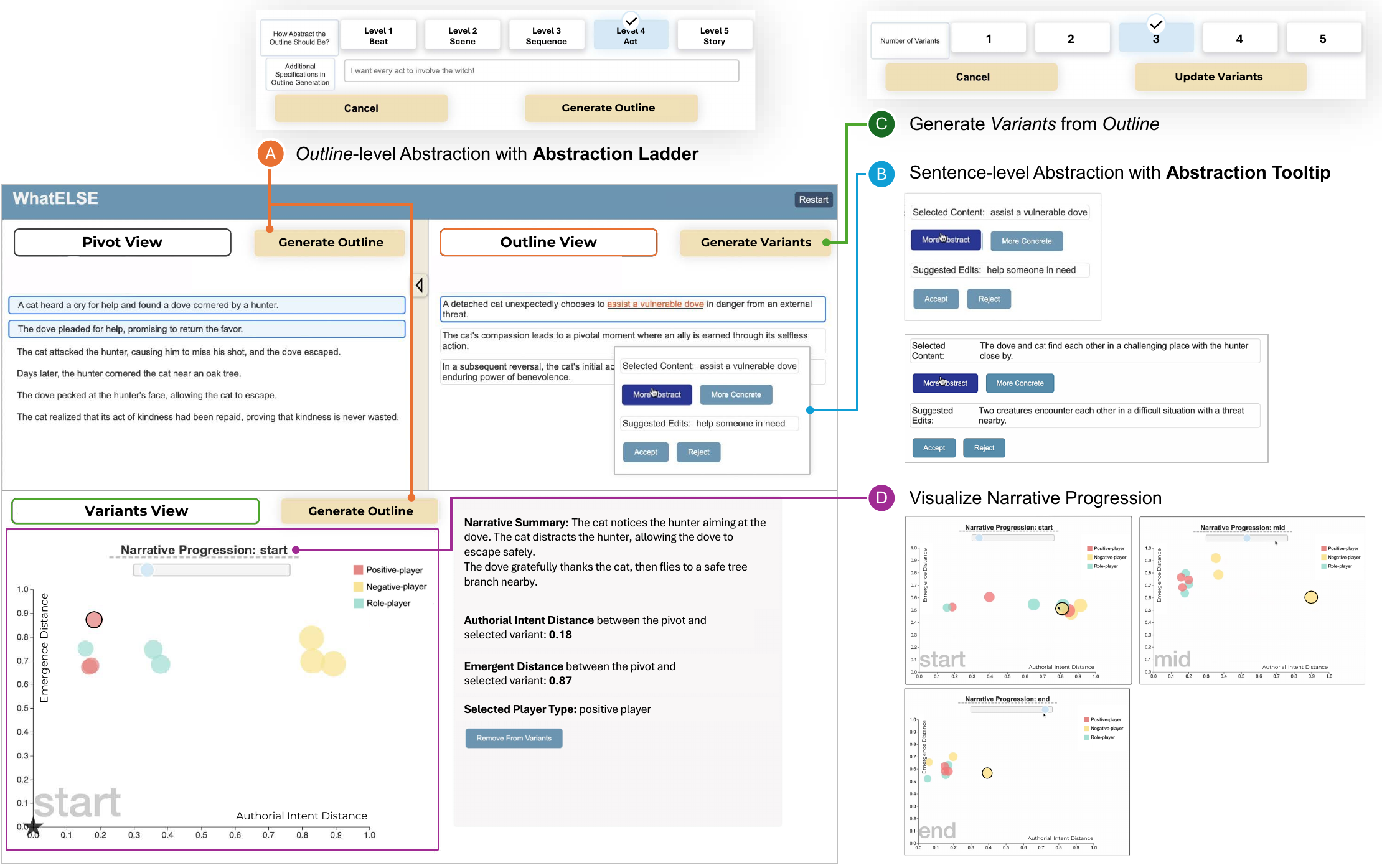}
\vspace{-10pt}
\caption{\presentation{An illustration of the Narrative Space Editor interface, including the pivot, outline, and variants view. Users can (A) generate outline from pivot or variants with an abstraction ladder to configure the abstraction level. They can (B) fine-tune sentence or word-level abstraction using an abstraction tooltip. They can also (C) generate variants from outline specifying the number of variants in the variants view. They can use (D) narrative progression slider to visualize the variants' dynamic distance from the pivot (star). }}
~\label{overview}
\vspace{-10pt}
\end{figure*}

{\sc WhatELSE} system assists the user in creating a narrative space. The user can upload narrative examples in text file(s). \presentation{In addition, the user uses a sentence to describe a story's moral (e.g., {\it ``kindness is never wasted''}). The system uses the story input to construct an initial version of the narrative space. The user can edit this narrative space using the interface.}
 {\sc WhatELSE} features three views for the user to perceive the narrative space: \textbf{Pivot View}, \textbf{Outline View}, and \textbf{Variants View} (Figure~\ref{overview}). 

\noindent \presentation{
\textbf{Pivot View}\hspace{1mm} The pivot view shows a pivot narrative instance. A {\em pivot (narrative instance)} is a user-defined narrative instance, considered as a representative instance in the narrative space. By default, the user's input is automatically marked as the pivot. The pivot serves as a point of reference as the user edits the narrative space.}

\noindent\textbf{Outline View} \presentation{An {\em outline} is an abstract specification of a sequence of events defining the narrative space. }\originality{Similarly to ``loglines'' \cite{mirowski2023co}, it specifically describes the general structure of the narrative with a sequence of high-level events }\presentation{ - e.g., {\it ``A small creature runs into an accident. It was then saved by another creature''}.} The outline describes the narrative space from a broader perspective by capturing the commonality across all the narrative instances in the narrative space. It represents the most abstract manifestation of the author's narrative intent, thus defining the boundary of the narrative space.

\noindent\textbf{Variants View}\hspace{1mm} A {\em variant (narrative instance)} is a narrative instance residing in the current narrative space. A variant instantiates the outline with a sequence of concrete events - e.g., {\it "An ant fell into water. A dove dropped a leaf next to the ant. The ant climbed on the leaf. The ant was saved."} Each abstract event in the outline is expanded to multiple concrete events in a variant.

The variants are displayed in an interactive scatter plot along two dimensions to help users understand the shape of the narrative space: 1) the \textbf{authorial intent} dimension, measured by the distance between the moral expressed by the variant and by the pivot (ranging from 0 to 1)\footnote{This distance is evaluated by prompting the LLM to assess how well the moral is conveyed}, and 2) the \textbf{emergence} dimension, measured by how much the plot progress in the variant deviates from the pivot (ranging from 0 to 1). \originality{These two dimensions are inspired by the ``authorial intent'' dimension in Riedl's taxonomy of IN approaches \cite{riedl2013interactive,riedl2009incorporating}, as well as the notion of ``emergence'' \cite{walsh2011emergent} and ``interactivity'' \cite{stang2019action} from prior IN research.} Users can configure the number of variants to be generated for visualization. Users can click on any variant in the visualization to display its detailed content, allowing them to compare it with the pivot. \presentation{Users can also use a scroll bar to visualize the plot progression} as they develop across different stages, allowing them to perceive how the narrative variants evolve over time and deviate from the pivot.

These three views provide different perspectives for users to inspect the narrative space. We also provide editing tools at each view to support shaping the narrative space in different ways. 


\subsubsection{Support Editing the Narrative Space} 

\presentation{The system provides editing tools at outline and instance level.}

\noindent\textbf{Outline Editing}\hspace{1mm} \presentation{Users can constrain or relax the boundary of the narrative space by adjusting the outline's level of abstraction.}  The more abstract the outline is, the less constrained the narrative space is. \presentation{For example, {\it ``a small creature got into an accident''} is more abstract than {\it ``the ant fell into water''}, enabling more possible narrative instances to be generated. The former removes the constraint on {\it ``the small creature''} being  {\it ``the ant''}, and the {\it ``accident''} being {\it ``falling into water''}.} A less constrained narrative space allows stronger player agency but follows a looser authorial structure. Outline editing allows the user to tune the narrative space to reach a desired balance between authorial structure and player agency. \presentation{The system provides two tools to support the abstraction editing}.

\begin{itemize}
    \item \textbf{Abstraction Ladder (Figure. \ref{overview}.A)} The abstraction ladder helps the user to shift the global level of abstraction across the events in the outline. Inspired by theories of narrative structure ~\cite{styan1960elements, mckee1997story}, \presentation{this ladder covers a spectrum of abstraction levels (beat, scene, sequence, act, and story level)}. An outline at the beat level is similar to a narrative instance, while an outline at the story level summarizes the plot into a one-line overview. Between the two ends, each level of abstraction is progressively more abstract than the previous level. For instance, a scene-level outline provides detailed descriptions of specific scenes, including characters, actions, objects, etc: \textit{``The kind dove takes a leaf to reach the ant and drags it out of a water bubble.''} An act-level outline offers a highly summarized view of the narrative, focusing on the turning points: \textit{``A character saves their friend from danger.''}
    \item \textbf{Abstraction Tooltip (Figure. \ref{overview}.B)} The abstraction tooltip \presentation{allows the user to adjust the sentence, phrase, or word-level abstraction in a more fine-grained manner.} Practically, when users select a text snippet in their outline plots, the tooltip appears, offering two options: ``More Abstract'' and ``More Concrete''. \presentation{By clicking the button, users receive suggested edits that replace the selected content with a more abstract or more concrete phrase.} While the abstraction ladder provides global control over the entire outline, the tooltip enables more fine-grained adjustments at the word or phrase level. The suggestion of making the selected content more abstract or more concrete is based on the taxonomy in linguistics~\cite{hayes1983cognitive}. For example, {\it ``character-animal-small animal-cat-tabby cat''} constructs a linguistic hierarchy. Given a selected text snippet {\it ``cat''}, requesting a more abstract suggestion would yield its superordinate term {\it ``small animal''} or {\it ``animal''}, while a more concrete suggestion would provide its subordinate {\it ``tabby cat''}. 
\end{itemize}
Once the user is satisfied with the outline, they can click the ``Generate Variants'' button to generate narrative variants in the Variant View. Section \ref{compiler} describes the technical pipeline for generating narrative instances from outline. 

\noindent\textbf{Instance Editing}\hspace{1mm} \presentation{Users can fine-tune the narrative space by editing the instance-level content in Pivot and Variant View. They can select a variant to set or unset it as the pivot. They can also remove a variant from the narrative space or add it back. Finally, they can directly edit the text in the instances. They can click the ``Generate Outline'' button to update the outline based on their edited variants. For example, a user who does not want to include certain player type may choose to remove all variants by that player type and update the outline. Section \ref{compiler} describes different player types in the player proxy model. }

\subsection{Technical Pipeline}


\label{sec:technical_pipeline}
This section describes our technical pipeline supporting the features described in the above section, focusing on the transformation between narrative outline and narrative instances. Specifically, we employ the GPT-4o ~\cite{openai2023chatgpt} for the implementation of our system.


\subsubsection{Transforming Narrative Instances to Outline} \label{summarizer}
We use an LLM prompting pipeline to generate outlines from narrative instances (Figure \ref{system_overview}.1). This pipeline first prompts the LLM with domain knowledge in drama writing, providing the context of the story domain and the narrative instances. The pipeline then prompts the LLM to summarize the commonalities across these narrative instances, generating outlines at different abstraction levels based on story structure principles \cite{mckee1997story}. Finally, the system selects an outline according to the user's chosen level of abstraction.

\subsubsection{Transforming Outline to Narrative Instances} \label{compiler}

\presentation{To generate meaningful events that can react to player actions (DG3), we go beyond text generation and integrate an LLM-based narrative planning approach with character simulation and player proxy models. Our method extends \textit{StoryVerse} \cite{wang2024storyverse} with player interactivity and behavior modeling.}
 \presentation{Generating narrative instances from outline is essentially simulating an interactive story generation process, where player actions may be generated by computational proxies of players, and the story generated grounded in the causal changes of game world states in accordance with the game mechanism.} 

\begin{figure*}[t]
\centering
\includegraphics[width=\textwidth]{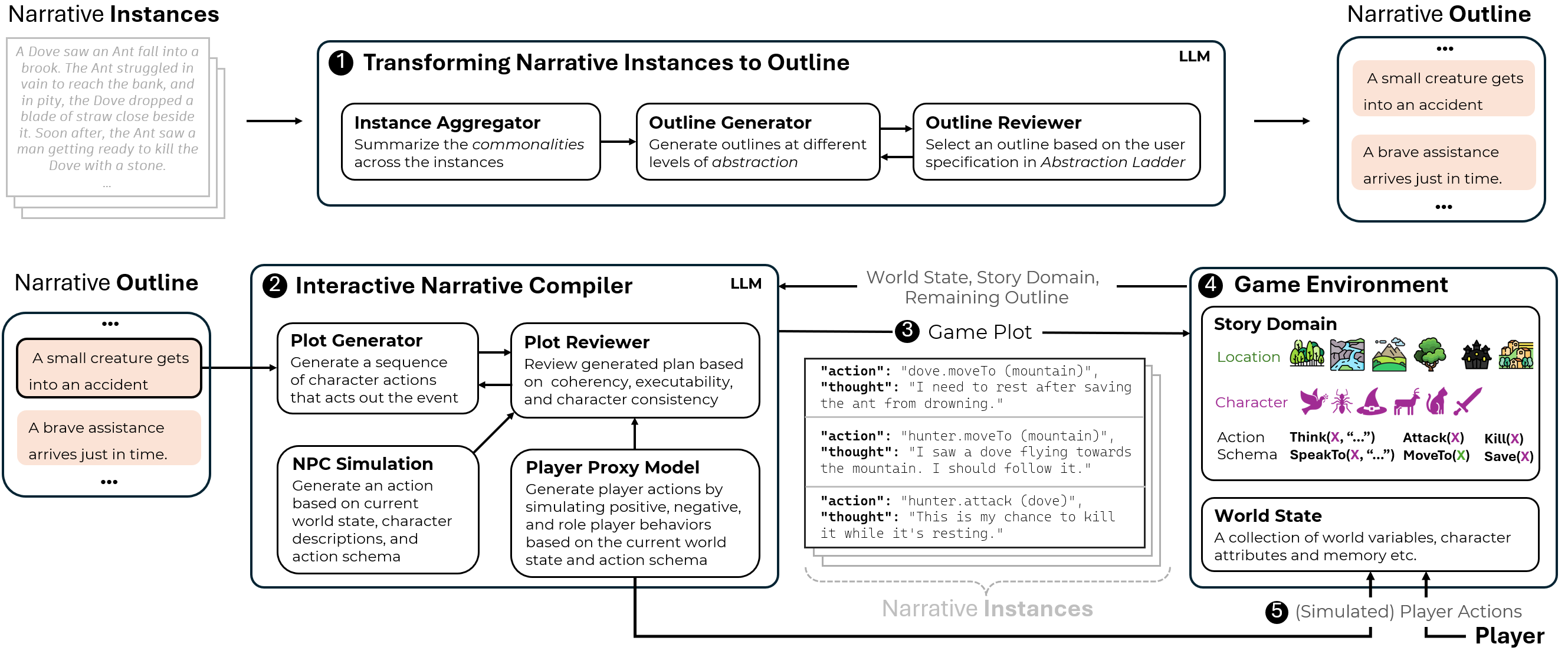}
\vspace{-10pt}
\caption{\presentation{An overview of the technical pipeline of \textsc{WhatELSE}. (1) The system transforms narrative instances to an outline using the LLM to summarize their commonalities, generate outlines at different levels of abstraction, and review the outline based on user specifications in the Abstraction Ladder. (2) The Interactive Narrative Compiler unfolds the outline into (3) a sequence of character actions to act out the events in the outline. (4) The Game Environment executes the actions and
updates the world states. (5) The player (or a simulated player) can interfere with the game by changing the world states. Finally, the Game Environment sends the updated world states and outline back to the compiler for the next iteration. }}
~\label{system_overview}
\vspace{-10pt}
\end{figure*}

To explain this process, we assume a {\em Game Environment} \presentation{(Figure~\ref{system_overview}.4) is given, which contains the {\em Story Domain} and maintains the {\em World State}.} The {\em World State} consists of a collection of variables that hold relevant values for the game mechanics, such as the characters’ attributes (e.g., health points), current locations, and relationship scores, as well as their memories from the simulation. \presentation{ The {\em Story Domain} contains a set of characters, locations, and an action schema that specifies executable actions in the game system. These actions are implemented as executable function calls that modify the variables of {\em World State} accordingly. For example, executing the action $\texttt{kill(X)}$ will result in character $\texttt{X}$'s state to become dead. }

\presentation{The main game loop starts by sending an event from the outline to the Interactive Narrative Compiler (Figure~\ref{system_overview}.2) to instantiate a sequence of character actions (Figure~\ref{system_overview}.3). The Game Environment (Figure~\ref{system_overview}.4) executes the actions and updates the world states resulting from the generated character actions. Once the Game Environment executes the actions, the player (or a simulated player) can interfere with the game by changing the world states, such as saving a character (Figure~\ref{system_overview}.5). Finally, the Game Environment sends the updated world states and outline back to IN Compiler for the next iteration.} The process loops over the events in the outline plot, and stops when it exhausts all the events.


\vspace{2mm} \presentation{\noindent \textbf{Plot Generator} \hspace{2mm} Given an event in the outline, the system generates a sequence of character actions that act out the event. It takes into account the current game world state as a result of all previous plot executions and player actions. }
An LLM is prompted to generate a sequence of character actions that act out the event. The prompt includes the following information from the game environment:
\begin{itemize}
   \item the list of characters and their descriptions;
   \item the action schema;
   \item current world state (including character's memory).
\end{itemize}

\presentation{This process is very similar to narrative planning which generates a sequence of state transitions that leads to a narrative goal. Compared to classic symbolic narrative planning, our narrative goal may be fuzzier - sometimes it may not be rigidly characterizable by world states. For example, the narrative goal could be {\it ``everyone likes Bob''}, which corresponds to world states semantically in a fuzzy way. This narrative goal can also be any arbitrary statements describing the desired outcome, constraining not only the endings but also the transitions, for example, {\it``someone was careless and got into an accident''}. Therefore, we use an LLM-based method instead of existing symbolic narrative planning methods for transforming outlines into concrete plots.}

\vspace{2mm} \presentation{\noindent \textbf{Plot Reviewer}} \hspace{2mm} \presentation{Similar to symbolic planning problems,  the plot generation requires causal soundness. This means the character actions must be valid state transitions according to the game's causal rules. We thus adopt an LLM-based planning method following the LLM-Modulo framework \cite{kambhampati2024llms}, with a game environment simulating plans generated by LLMs and providing external critiques.}
To review the generated plan, an LLM is prompted to provide feedback regarding the quality and feasibility of the action sequence to improve it:
    \begin{itemize}
    \item \textbf{Overall Coherency Evaluation} Feedback is obtained by prompting an LLM to comment on the overall coherency of the generated plot and make suggestions for improvement.
    \item \textbf{Character Simulation Evaluation} For every action in the sequence, we prompt an LLM to play the role of the subject of the action. Given the current world state including the character’s memory, we ask the LLM if the motivation for the character to perform the action has been established. We include the explanation to this question in the feedback if the motivation has not been established.
    \end{itemize}
    In addition, we leverage a simulated Game Environment for external evaluation. The system evaluates the \textbf{Causal soundness} of the generated action sequence and reports the observations on the success/failure of the execution. The combined feedback is added to the prompt for the next round of generation.

\vspace{2mm} \presentation{For example, the event {\it ``a small creature gets into an accident''} could be turned into a sequence of character actions shown in Figure~\ref{system_overview}.3. Note that the events in the outline plot are at a higher abstraction level. This means the same event can be transformed into multiple character action sequences, leaving room for the exact plot to adapt to different world states \footnote{In the above example, if the dove is dead at the time of plot execution, then a different character action sequence will be generated - one possibility is that the ant fell into the water.}.
Once the final sequence of character actions is generated, it will be executed by the {\em Game Environment} to update the world state. }

\presentation{The Plot Generator and Reviewer create a sequence of character actions to act out the event. In between these events, the player or NPCs take free actions. These actions are driven by the LLM.
The player actions are determined \presentation{either by a real player's input or a simulated Player Proxy Model (Figure~\ref{system_overview}.5)}. }

\vspace{2mm}\noindent \presentation{\textbf{Player Proxy Model}} \hspace{2mm} 
When generating narrative variants, player actions are provided by an LLM-based player proxy model which operates based on player behavior classification derived from previous studies in digital games ~\cite{yannakakis2013player,worth2015dimensions}. Our player simulation incorporates three key player behavior models:

\begin{itemize}
    \item \textbf{Positive Players} in digital games contribute positively by following the intended game objectives and exhibit helping behaviors~\cite{velez2013helping,bostan2009player}.
    \item \textbf{Negative Players} are the killers identified in classic player modeling~\cite{majors2021some,hamari2014player}. They often exhibit aggressive behavior that disrupts the experience of others, particularly when they seek to dominate or harm others destructively.
    \item \textbf{Role Players} prioritize narrative immersion and character development by mimicking the actions their character would take in the gaming world~\cite{praetorius2020avatars}.    
\end{itemize}

Using these player models, we simulate a potential plot that could emerge from the interaction between game characters and simulated players within the narrative space defined by the outline. In this way, the system generates a diverse set of narrative instances in the variants view.

\vspace{2mm} \noindent \presentation{\textbf{Non-Player Character Simulation}} \hspace{2mm} 
An LLM is prompted to role-play as each of the NPCs and generate an action for this character. The prompt includes the following information:
\begin{itemize}
   \item the action schema;
   \item the list of characters and their descriptions;
   \item current world state (including character's memory);
\end{itemize}
Note that the character actions are not directly determined by the outline. However, it is affected by the current world state and, therefore, indirectly influenced by the executed events in the outline.

\vspace{2mm}
\presentation{Using this pipeline, \textsc{WhatELSE} creates a gameplay experience by unfolding the outline into narrative instances. The system generates the game plot for each event in the outline as a series of character actions. The player then inputs actions within the action schema, which influence the progression of the subsequent plot. The system runs executable actions to update the world state. After each round of player action, the system unfolds the next events until exhausting all the events in the outline, in this way, creating an interactive narrative experience. }

\input{sections/exampleWorkflow}

%% file: sections/exampleWorkflow.tex
\subsection{Example Workflow}


Below we present an example workflow to demonstrate some of the features described above. 
\usecase{Alice, a novice text-adventure game designer, wants to create a game based on the setting of a novel she enjoys.} 
Alice opens \textsc{WhatELSE}, along with a game engine preloaded with a story domain based on the novel.

\subsubsection{Encode Authorial Intent in Narrative Space} 
\begin{figure*}[h]
\centering
\includegraphics[width=1.0\textwidth]{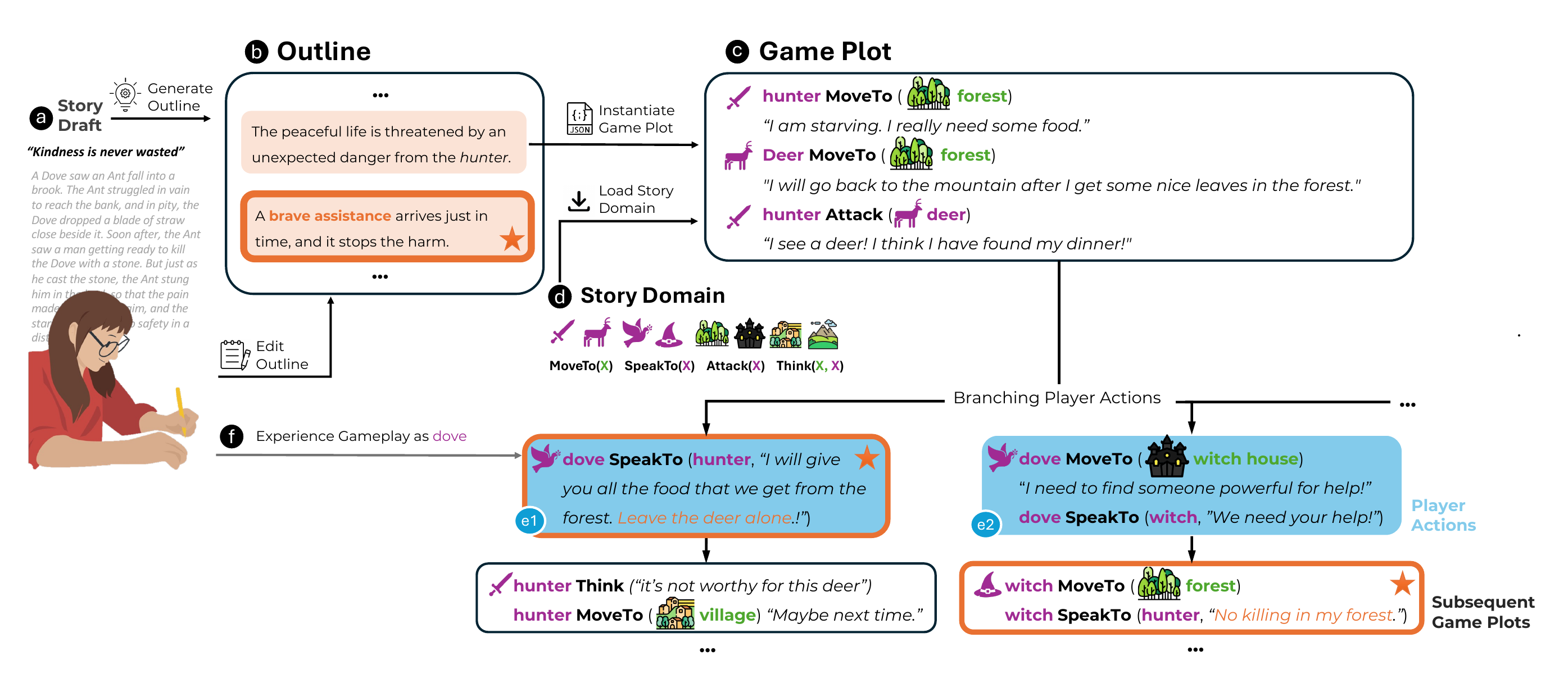}
\vspace{-20pt}
\caption{\presentation{An example workflow that shows (a) an author uploads a story draft in \textsc{WhatELSE} to (b) generate an outline. The system unfolds the outline into (c) an executable game plot with (d) a pre-loaded story domain, which supports branching storylines based on the player actions. If the player chooses to (e1) save the deer from the hunter, this action fulfills the ``brave assistance'' event in the outline defined by the author (shown as the orange star). If the player chooses to (e2) ask another character (e.g. a witch) for help, the witch will instead save the deer, demonstrating ``brave assistance'' to fulfill the event. Alternatively, if the player does not choose to save the deer at all, the system will choose a character from the story domain to save the deer as a demonstration of ``brave assistance''. This example shows how the game plot is dynamically adjusted based on the player actions to fulfill the outline. (f) The author can play the game plot to better understand the player experience. }}
~\label{walkthrough}
\vspace{-10pt}
\end{figure*}

Alice starts with a rough draft of the story and a moral she wants to convey: {\it ``Kindness is never wasted''}. Using \textsc{WhatELSE}, she uploads her initial story into the system \presentation{(Figure~\ref{walkthrough}.a)}. The story is displayed in the pivot view, showing a sequence of events; while an initial outline appears on the right, summarizing the key turning points \presentation{(Figure~\ref{walkthrough}.b)}. Alice adds details to the pivot to refine her story. Once satisfied, she clicks the Generate Outline button to update the outline based on her edits. She chooses the ``act level'' and specifies, {\it ``The hunter has to appear in every act''}. Alice hovers to see how each event in the outline is mapped to the entries in the pivot plot. She continues exploring different levels to find the ideal level of abstraction.

\presentation{Alice finds one of the events ({\it ``The peaceful life is threatened by an unexpected danger from the hunter''}) to be too restrictive for the hunter to cause the danger. She uses the abstraction tooltip to replace the phrase {\it ``the hunter''} with {\it ``a character''} to leave room for variations in the game.} 
Alice looks at the outline and is unsure what players might experience. So she clicks the Generate Variants button. The interface displays a scatter plot of potential narrative instances. Alice scrolls through different plot stages of these instances — from start to end — she notices that some instances continuously express the moral, while others only reveal it toward the end, both of which she considers acceptable. However, she also spots a cluster of instances that fail to express the moral by the end of the narrative. Curiously, she clicks on a dot representing one of these instances and reviews its details. 

Alice reads the instance and realizes the issue is in the event that she had previously set as {\it ``a character''}, which was too loosely defined, allowing the system to choose an undesirable character. To address this issue, she changed it back to {\it ``a human character with power''}, allowing the system to choose a character reasonable for the second event. 

Later, Alice notices a set of three variants where one of the events unfolds as, {\it ``the dove speaks with the hunter, leading the hunter to notice and then chase the dove''}. Alice finds this version more compelling than her pivot plot. She removes other variants, only leaving these three narrative variants in the view. Satisfied with these variants, she clicks the Generate Outline button to create a new outline that summarizes their commonalities. She then returns to the outline editor, using the abstraction tools to iteratively edit the outline, until it aligns with the story's moral and represents a narrative space that incorporates the interesting variations.

\subsubsection{Unfold the Narrative Space For interactivity}
With the narrative space defined by the outline, \usecase{Alice can experience the narrative instances unfolding in a turn-based text adventure game. She goes to the interactivity page. The system loads the story domain that includes a set of characters, locations, and action schema \presentation{(Figure~\ref{walkthrough}.d)}. The first sequence of the game plot is generated: a hunter is looking for food and finds a deer to hunt \presentation{(Figure~\ref{walkthrough}.c)}}. 

Alice, playing as the dove, chooses her next moves from a pin pad \usecase{\presentation{(Figure~\ref{walkthrough}.f)}. She can bravely stop the hunter by giving out her food \presentation{(Figure~\ref{walkthrough}.e1)}. Alternatively, she could ask other characters for help \presentation{(Figure~\ref{walkthrough}.e2)}. 
The system compares the player's action with the narrative outline. If the player chooses to save the deer on their own \presentation{(Figure~\ref{walkthrough}.e1)}, the event of {\it ``brave assistance''} is fulfilled by the player action. If the player chooses other actions, the system will create an event where another character demonstrates {\it ``brave assistance''} \presentation{(the witch in Figure~\ref{walkthrough})} to fulfill the event. The system generates subsequent character behaviors based on the player action. }

This turn-based interaction continues, with Alice alternating between reviewing generated game plots, observing character simulations, and experiencing the generated game play as a player. \usecase{Since she wrote a total of five events in her outline, the game play proceeds for five rounds, until all the events she planned have been played out. Since the game plots generated are fully structured, Alice can directly export the output of the narrative compiler as a finite state machine into the game engine where she can visualize the characters and locations.}

\usecase{
\subsubsection{Additional Use Cases} In addition to Alice's case as a text-adventure game designer, \textsc{WhatELSE} can also serve as a powerful tool for a wide range of users. Game masters, mod developers, and fan creators across different domains can leverage its capabilities. For example, dungeon masters in tabletop role-playing games can use the Narrative Space Editor to outline gameplay scenarios before sessions and employ the Interactive Narrative Compiler to dynamically determine outcomes of player actions during gameplay. Fan creators~\cite{booth2009narractivity} can efficiently transform their favorite novels, movies, or other media into interactive narratives, using the \textsc{WhatELSE} to structure and unfold new, personalized storylines based on the original story domain. Beyond entertainment, educators can utilize \textsc{WhatELSE} to design interactive learning experiences, such as gamified learning tutorials or interactive training modules. }

%% file: sections/userStudy.tex
\label{sec:user_study}

We conducted a user study with 12 participants to evaluate {\sc WhatELSE}. The goals of the study were to (1) understand their usage and perception of the Narrative Space Editor, (2) validate the effectiveness of Interactive Narrative Compiler in preserving authorial intent, and (3) explore the benefits and drawbacks of the AI-bridged IN creation workflow.

\subsection{Participants}
We recruited 12 participants (6 male, 6 female, between 20 to 34 years old, average 27 years) from within our institution \evaluation{by posting on an internal network channel. The demographic information of the participants was provided in the supplementary material.}
Participants were asked to self-report their experience in using generative AI (e.g., ChatGPT), digital games, and interactive narratives. Each was compensated with a \$75 gift card. All participants reported at least moderate experience in using generative AI. More than half of the participants used to experience interactive narratives. Five participants reported spending more than two hours a week playing digital games. 
Nine participants reported having no experience creating interactive narratives, while the other three reported having very limited experience.

\subsection{Task}
\begin{figure*}[h]
\centering
\includegraphics[width=\textwidth]{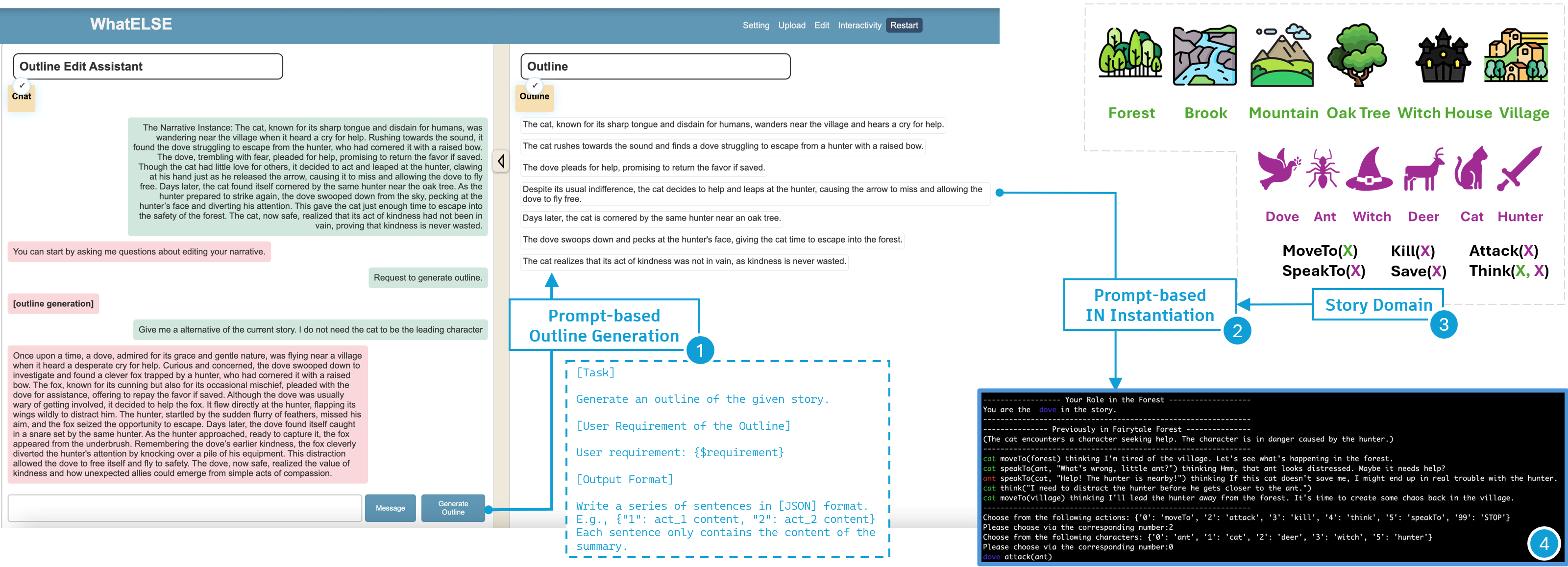}
\vspace{-10pt}
\caption{\baseline{An illustration of the baseline system that uses a prompt-based approach for (1) the outline generation and (2) IN instantiation with (3) the pre-loaded Fairytale Forest story domain. (4) During runtime, LLM acts as a text adventure game engine, dynamically generating game plots based on the outline and responding to player input in the baseline system. }}
~\label{baseline}
\vspace{-10pt}
\end{figure*}

The task is to create a playable game plot using the systems provided.
Given a story domain and a narrative example with a story moral, participants were asked to generate an outline to guide the creation of a game plot that expresses the same moral as the given narrative instance.
The task uses the \presentation{\textit{"Fairytale Forest"} story domain (Figure~\ref{baseline}.3)}. It contains six characters (ant, dove, hunter, witch, cat, and deer), five locations (mountain, forest, village, brook, and witch house), and an action schema with six actions ("move to", "speak to", "kill", "attack", "think", "save") that defines what characters can perform. 
Based on the story domain, we adapted two stories widely used in previous studies as narrative examples  ~\cite{suwartini2019culture,affendi2018elements}. These stories feature clear plots centered on emerging danger and the characters' decisions to help each other. The two narratives convey two distinct morals: \textit{``Kindness is never wasted''} and \textit{``Kindness is not always rewarded''}. Details of story domain and examples can be found in the supplementary material. We chose this story domain and narrative examples because they are simple and easy to follow, making it suitable for novice users with limited experience in IN. We used the pre-defined story moral as a representation of the authorial intent.

To complete the task, participants are encouraged to meet these criteria for their final outline: (1) The outline should consist of 3–4 sentences. (2) Each sentence must have at least one abstract element, allowing multiple events to fit the description. (3) The outline should convey the same moral as the provided narrative instance. (4) Participants are encouraged to make creative edits to the narrative instance, ensuring that their authorial intent is reflected in the outline. These criteria guide participants by providing clear objectives and controlling the process.

\subsection{Conditions}

\baseline{We compared \textsc{WhatELSE} with a baseline system that uses a conversational LLM-powered assistant in a counterbalanced within-subject design. This comparison aimed to evaluate the effectiveness of \textsc{WhatELSE} in supporting two features: (1) generating outlines from a narrative instance and (2) instantiating narratives from the outline (IN Compiler). Specifically, neither of the features is supported by traditional IN authoring tools such as Twine \cite{friedhoff2013untangling}. They are also not natively supported by a plain ChatGPT-like interface. Therefore, we implemented a baseline system with these two features empowered by LLMs for a fair comparison. 
Participants could load a narrative instance and use the chatbox with a send button to prompt natural language queries. By pressing the ``Generate Outline'' button, an auto-generated outline would be displayed on the right. This outline generation used a prompt-based approach that loaded the chatbox history as requirements and outputted the outline in JSON format (Figure~\ref{baseline}.1). Participants can also directly edit the generated outline. To instantiate narratives from the outline, we implemented a prompt-based plot generation approach (Figure~\ref{baseline}.2, prompt in Appendix) similar to prior work \cite{ai_dungeon2}, with preloaded knowledge of the story domain on characters, locations and action schema (Figure~\ref{baseline}.3). During runtime, the LLM acts as a text adventure game engine, dynamically generating game plots based on the outline and responding to player input in the baseline system (Figure~\ref{baseline}.4). }

\subsection{Procedure} 

All 12 studies were hosted in person. After obtaining consent, an experimenter introduced the background of IN, including the concepts of narrative instance, narrative space, and playable game plots. \evaluation{Participants were oriented to complete the task by the experimenter, who provided instructions, answered questions, and offered help throughout the study}. They were informed of two systems. Each system has two stages: creating an outline and playing the game plot generated from it. 

In each condition, participants had five minutes to prepare a narrative instance with a story moral. An example was provided as a starting point. To prime them for a creator's mindset, participants were encouraged to edit the given story by highlighting interesting scenes or adding details, ensuring their edits still aligned with the story moral.
Upon editing the story, participants received a walk-through tutorial on the assigned system's features. Then, they had five minutes to practice and familiarize themselves with the tool. 
After the tutorial, participants used the assigned system to create a narrative outline for 15 minutes. They were encouraged to iteratively refine their outline until it met the task criteria. Upon completion, participants were asked to fill out a narrative editing questionnaire (Figure~\ref{fig:questionnaire}.a) rating their experience with the system. 

After filling out the questionnaire, participants played a turn-based interactive narrative game as the Dove in the Fairytale Forest. In each turn, they could choose their actions, with the plot developments guided by the outline they created. They then filled out a game plot questionnaire (Figure~\ref{fig:questionnaire}.b) rating their experience with the generated plots.  Finally, we conducted a 10-minute post-study semi-structured interview about their preferences for the two systems. The study took about 70 minutes in total.


\evaluation{
\subsection{Measurement and Analysis}
We collected answers from a semi-structured interview and two questionnaires (narrative editing and game plot questionnaire). We also recorded their interaction activity in the interface. 
The narrative editing questionnaire focused on assessing the authoring experience. It has 11 questions, including 7 questions rating the controllability, expressivity, and perceived overall experience of the system, as well as 4 questions from the NASA Task Load Index (TLX) ~\cite{hart1988development} which measured participants' perceived workload while using the system. 
The game plot questionnaire focused on evaluating the player experience and their perceived quality of the generated game plots. This questionnaire included 9 questions rating controllability, expressivity, and the overall quality of the game plots.
To examine the potential differences between conditions, we conducted the Wilcoxon signed rank test on the questionnaire results. For the qualitative data collected in the interview, we transcribed audio from the interview and analyzed the results using thematic clustering ~\cite{blandford2016qualitative}. 
}

\subsection{Results}
\begin{figure*}[t]
  \centering
  \subfloat[Outline Editing Questionnaire - Experience and Perceived Quality]{\includegraphics[width=0.98\textwidth]{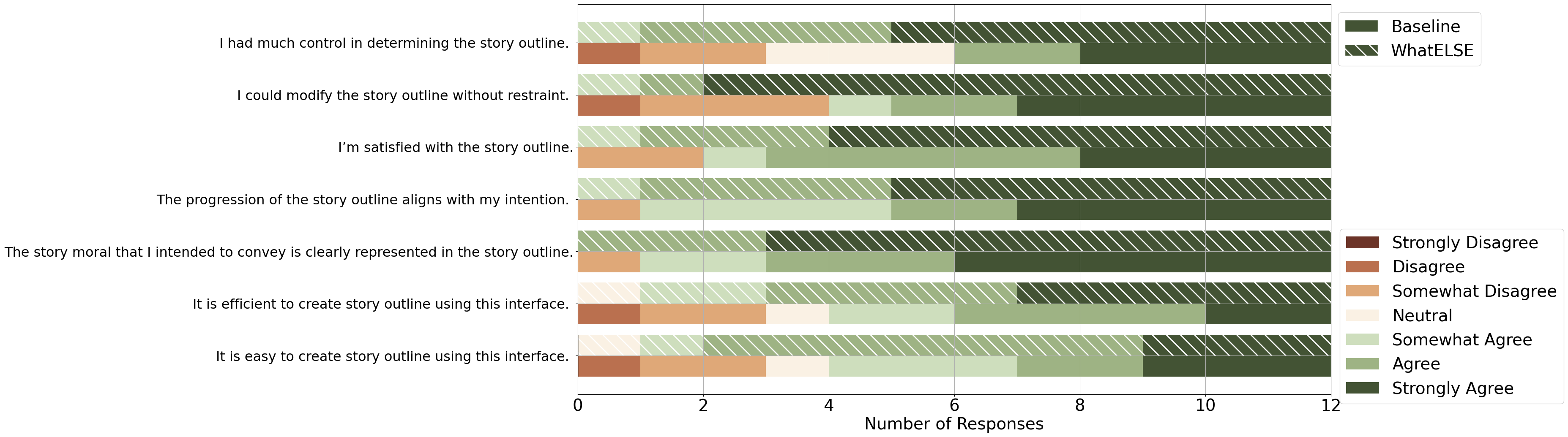}\label{user_study:interface}}
  \hfill
  \subfloat[\evaluation{Outline Editing Questionnaire - NASA TLX Rating.}]{\includegraphics[width=0.9\textwidth]{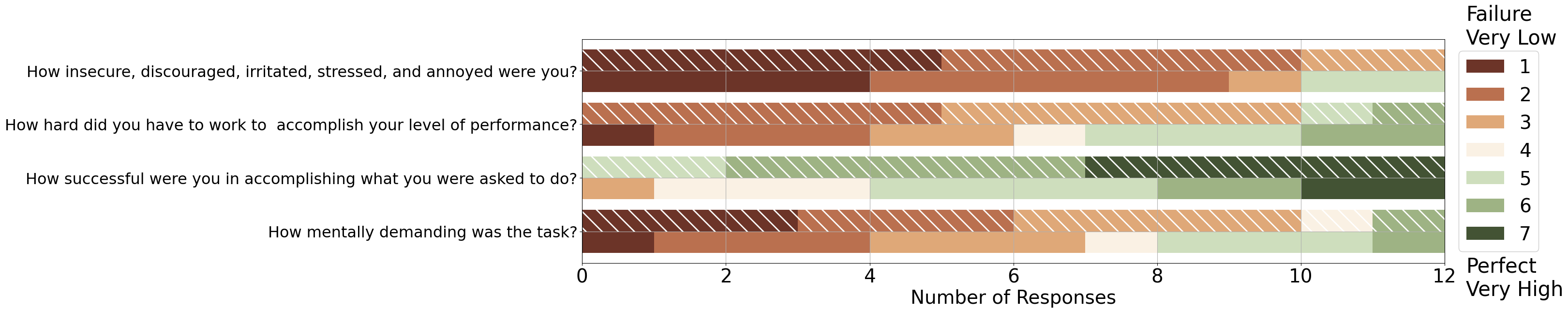}}\label{user_study:interface}
  \hfill
  \subfloat[Game Plot Questionnaire]{\includegraphics[width=0.98\textwidth]{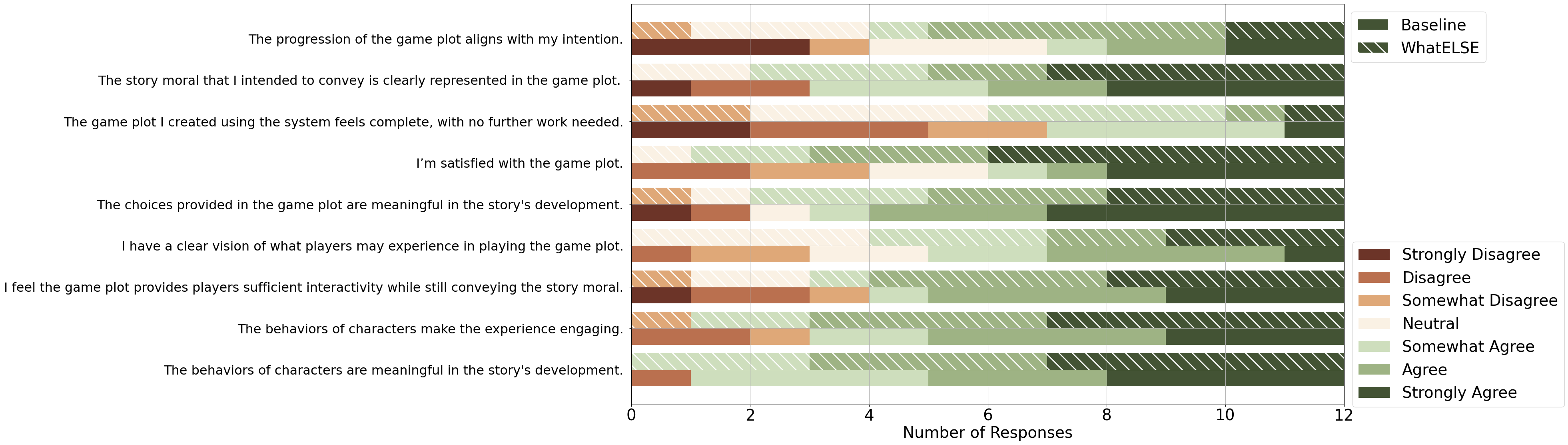}\label{user_study:game_plot}}
  \vspace{-5pt}
  \caption{Questionnaire results comparing \textsc{WhatELSE} with the baseline.}
\label{fig:questionnaire}
\vspace{-10pt}
\end{figure*}

\begin{figure}
    \centering
    \includegraphics[width=.95\columnwidth]{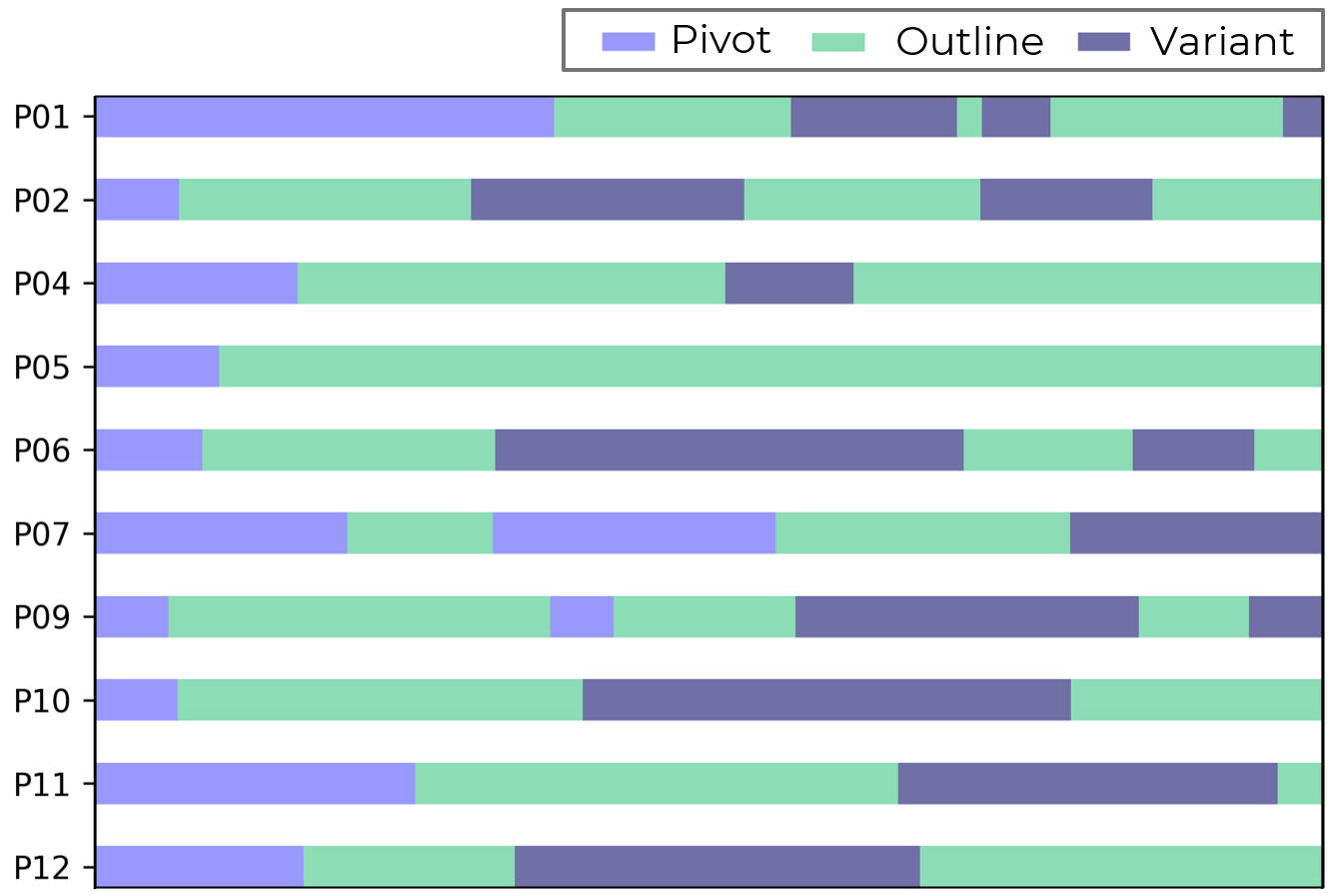}
    \caption{Percentage of participants' interaction time in each view using \textsc{WhatELSE} to create an executable interactive narrative.}
    \label{fig:behaviors}
    \Description{}
\end{figure}

We present questionnaire results (Figure~\ref{fig:questionnaire}), observation of participants' interactions\footnote{A recording error resulted in the loss of two participant’s interaction recordings.} (Figure~\ref{fig:behaviors}), and their interview responses to understand how \textsc{WhatELSE} helps people shape the narrative space and further unfold it into executable game events. Numbers in brackets indicate Median ($Med$).  

\paragraph{WhatELSE supports the creation of a narrative outline through authorial controls.} Participants reported to have more control in generating the outline using \textsc{WhatELSE} ($Med$: 7) than baseline ($Med$: 5, $p = 0.048$). They also reported that they were less restrained in creation using \textsc{WhatELSE} ($Med$: 7) than the baseline ($Med$: 6, $p = 0.049$). In the interview, most participants (9 out of 12) commented that the abstraction tools were helpful. As P4 noted, the ability to select the level of abstraction \textit{``allows users to have more control via clicking buttons.''} Similarly, P7 highlighted that the Abstraction Tooltip enhanced the creation process by providing \textit{``an easier interface to change anything you got."} Overall, participants were more satisfied with the outline generated using \textsc{WhatELSE} ($Med$: 7) compared to the baseline ($Med$: 6, $p = 0.047$). They felt more successful in completing the task of creating outlines with \textsc{WhatELSE} ($Med$: 6) compared to the baseline ($Med$: 5, $p = 0.025$).


\paragraph{WhatELSE helps participants perceive the Narrative Space} We found participants alternated between the narrative outline and narrative instance to perceive the narrative space. Nine participants alternated between Variant and Outline View to edit the narrative space, and two participants alternated between Pivot and Outline View (Figure~\ref{fig:behaviors}). While the Outline View gives a rough idea of the narrative space, participants found the Variant View offered more detailed information. For example, P2 found the Outline View useful but also commented it \textit{``very summarized though''}. They then switched to the Variant View for concrete examples, which provided them with \textit{``interesting variations''} (P2) that they found \textit{``very helpful for inspiration''}. We observed multiple participants having similar surprising reactions to the variants, indicating that people may \textit{underestimate} the scale of narrative space by working on the outline. Participants also used the Variants View to prevent deviation from their authorial intent. For example, P1 frequently checked the variants for their distance to the given story moral, stating, \textit{``I pretty much completely relied on that, right? OK, if anyone here (character) isn't following my vision for the story, then, you know, I can easily know what the problem is.''} This indicates that the visualization helped him quickly identify unexpected elements in the narrative space. 


\paragraph{WhatELSE preserves the authorial intent in the gameplay. } While playing the game generated from the outline, participants reported that the progression of the game plot aligns better with the story moral using \textsc{WhatELSE} ($Med$: 6.0) compared to the baseline ($Med$: 4.0, $p = 0.020$). For instance, P10 expressed that in the baseline condition, the story moral in the game felt \textit{"some kind of lost"}, indicating a misalignment with their expectations. This contrasts with P10's experience in WhatELSE, where the moral was better preserved within the plot. P5's experience also showcases the effectiveness of the Interactive Narrative Compiler in preserving the authorial intent. P5 created an outline to express \textit{``kindness is not always rewarded''}. In the first round of gameplay, they killed both a weak character (an ant) and a threatening hunter instead of saving the ant. In the second round, the system adapted by having other characters (a deer and a cat) try to stop P5's killing behaviors, but P5 killed them too. In the final round, the system output a monologue that forced P5 to reflect on their actions, conveying the lesson that \textit{"kindness is not always rewarded"}, with kindness coming from NPCs rather than from the players. P5 praised \textsc{WhatELSE} for its adaptability and for effectively conveying the intended moral, describing the system as \textit{``smart''} for how it \textit{``tied my actions and packaged it to be its original directive''}, which was the story moral.

\paragraph{WhatELSE provides engaging game events reacting to player actions. } Overall, participants were more satisfied with the game events generated by \textsc{WhatELSE} ($Med$: 6.5) over the baseline ($Med$: 4.5, $p = 0.014$). They felt that \textsc{WhatELSE} responded significantly to their actions, enhancing their sense of agency. For instance, in the baseline, P6 killed a character, but due to a lack of narrative planning, the supposedly dead character reappeared after several rounds. \presentation{However, in {\sc WhatELSE}-generated games, their actions had a greater impact}; when they tried to "kill" a character, the game responded appropriately. They also found the game plot to be more complete and they will spend less further work in editing narrative space ($Med$: 4.5) compared to the baseline ($Med$: 3.0, $p = 0.027$).

%% file: sections/technicalEvaluation.tex
The user study shows that \textsc{WhatELSE} effectively helps users create engaging interactive narratives by enhancing both authorial control and player engagement through efficient narrative space editing. To validate the technical pipeline driving the transformation between narrative outline and instances, we conducted technical evaluations focusing on effectiveness in 1) generating the outline from instances and 2) generating instances from an outline.

\subsection{From Narrative Instances to Narrative Outline}

\presentation{\textsc{WhatELSE} provides an abstraction ladder with different levels of abstraction to generate an outline from instances using a prompting pipeline.} To examine the effectiveness of the pipeline, we employ two lexical-level measures, {\em concreteness rate}, and {\em imageability score}.  Both measures are adapted from large-scale crowdsourced annotations in previous studies that have been widely used in linguistic evaluations ~\cite{wilson1988mrc,brysbaert2014concreteness}. \evaluation{Both scores are lexicon-based, with each word assigned an averaged score from a batch of crowdsourced annotations; for each outline, we calculate the average score across all words in this outline, excluding stop words.} Intuitively, a higher concreteness rate indicates that the wording is more concrete and specific, corresponding to a lower level of abstraction. A lower imageability score suggests that the wording allows greater room for interpretation, corresponding to a higher level of abstraction. 

\presentation{We generated outlines for a sample of 100 stories from the Fairytale dataset~\cite{xu2022fantastic} at three different levels of abstraction (scene, sequence, and act level).} We reported the concreteness rate and imageability score of the generated outlines in Table ~\ref{tech_eval:abstraction_ladder}. The results show that our method effectively produces outlines at three distinct levels of abstraction, with significant differences in the lexical measures between each pair of abstraction levels. This demonstrates the system's ability to define narrative spaces with varying degrees of constraint.



\begin{table}[h]

\small
\begin{tabular}{c|c|c|c}
\toprule

\multirow{2}{*}{\textbf{Measurement}}
& \multicolumn{3}{c}{\textbf{Abstraction Level }} 

\\
\cline{2-4}
 & \textbf{Scene Level}  & \textbf{Sequence Level} & \textbf{Act Level}
\\
\midrule

Concreteness Rate & $3.56\pm 0.05$ & $3.09\pm0.06$  & $2.95\pm 0.06$ 
\\


Imageability Score & $497.33\pm{22.82}$ & $453.64\pm{32.42}$ & $438.91\pm{34.17}$
\\

\bottomrule
\end{tabular}
\vspace{2pt}
\caption{Measured abstraction level of the generated outline plot using the proposed prompt pipeline employed in Abstraction Ladder, to generate outline with distinct levels of abstraction from narrative examples. A lower concreteness rate and imageability score indicate the text is more abstract at the lexical level.}
\label{tech_eval:abstraction_ladder}
\end{table}




\begin{table*}[ht]

\begin{tabular}{c|c|c|c|c}
\toprule

\multirow{2}{*}{\textbf{Measurement}}
& \multicolumn{2}{c|}{\textbf{Human-generated Outline}} & \multicolumn{2}{c}{\textbf{LLM-generated Outline}} 

\\
\cline{2-5}
 & \textbf{$d_1$}  & \textbf{$d_{macro}$} & \textbf{$d_1$}  & \textbf{$d_{macro}$}
\\
\midrule


\presentation{Proposed IN Compiler} & {$0.65\pm 0.01$}             & $0.78\pm 0.01$  & $0.64\pm 0.01$ & $0.77\pm 0.01$
\\

\presentation{Baseline Prompt-based IN Compiler} & $0.51\pm 0.02$             & $0.63\pm 0.02$  & $0.61\pm 0.02$ & $0.73\pm 0.02$
\\

\bottomrule
\end{tabular}
\caption{Measured plot distance by averaged ROUGE-1 distance $d_{1}$ and the macro-averaged ROUGE distance $d_{macro}$ among the plots generated by baseline and our approach. Results show that \textsc{WhatELSE} IN Compiler leads to a larger averaged distance among the plots, and thus a greater diversity of plots within the narrative space.}
\label{tech_eval:plot_diversity}
\end{table*}

\begin{table*}[h]

\begin{tabular}{c|c|c|c|c}
\toprule

\multirow{2}{*}{\textbf{Method}}
& \multicolumn{4}{c}{\textbf{Measurement}} 

\\
\cline{2-5}
 & \textbf{$d_1$}  & \textbf{$d_{macro}$} & {World-state Change (Neg)}  & {Character Involvement (Pos)}
\\
\midrule


\presentation{Proposed IN Compiler} & {$0.59\pm 0.01$}             & $0.74\pm 0.001$  & $1.00\pm 0.0$ & $1.70\pm 0.37$
\\

\presentation{Baseline Prompt-based IN Compiler} & $0.53\pm 0.01$             & $0.67\pm 0.01$  & $0.85\pm 0.08$ & $1.65\pm 0.41$ 
\\

\bottomrule
\end{tabular}
\caption{Measured impact of player action on game plot progression, by (1) averaged pairwise ROUGE-1 distance $d_{1}$ and the macro-averaged ROUGE distance $d_{macro}$ between pairs of the game plots driven by contrasting player actions, and (2) averaged world state change rate driven by the negative player action and the averaged character involvement driven by the positive action. Results show that the contrasting player actions make the proposed approach generate plots with larger pairwise distances. Additionally, in \textsc{WhatELSE} playtime, player actions lead to more stable world-state change and better character involvement than the baseline. }
\label{tech_eval:action_impact}
\end{table*}

\subsection{From Narrative Outline to Narrative Instances}
\presentation{We compared the plot quality generated using our approach with the baseline prompting-based approach (Figure~\ref{baseline})} from two aspects: {\em plot diversity} and {\em player impact}.

\noindent \textbf{Plot Diversity}\hspace{2mm}  Plot diversity refers to the ability to generate a wide range of different plots within the narrative space. It indicates the level of interactivity and player agency supported by the plot generation method, as it showcases the system's ability to offer varied storylines within the narrative space described by the outline.

To quantitatively assess diversity among a set of $N$ plots generated within the narrative space, we calculate the averaged distance between each plot and the other $N - 1$ plots in the set. We then compute the average distance for each plot relative to the others. For distance calculation, we use the ROUGE score ~\cite{lin2004rouge}, a reference-based evaluation metric that measures text similarity. Adapting from the ROUGE-1 score $r_{1}$ targeting word level and macro averaged ROUGE score $r_{macro}$ measures similarity across multiple levels of wording, we compute the word-level distance $d_{1} = 1 - r_{1}$ and the macro distance $d_{macro} = (1 - r_{macro})$ between plots, respectively. As the ROUGE score indicates similarity between text, a higher averaged distance indicates the greater diversity of plots.

For comparison, we first collected two sets of outlines based on the \presentation{\textit{``Fairytale Forest''} story domain (Figure~\ref{baseline})}. The first set, consisting of 12 outlines, was generated by participants in the user study, each tied to one of two specific morals. Additionally, we developed a set of 50 outlines by prompting an LLM, focusing on various morals within the story domain. We simplify each outline by taking only the first act, resulting in 12 human-generated and 100 LLM-generated single-act outlines. These outlines were then used to guide plot generation without involving player actions, using the proposed narrative planning based approach and the baseline prompt-based approach.

We generated a set of 20 plots with each outline and then calculated the averaged distances $d_{1}$ and $d_{macro}$ among each set of plots. As shown in Table~ \ref{tech_eval:plot_diversity}, our approach generated more diverse plots within the same narrative space, indicating more variety of storylines and stronger player agency.


\noindent \textbf{Player Impact} \presentation{We use the term player impact to refer to the extent to which players' actions meaningfully influence the progression of the plot.} A higher player impact indicates that the narrative is more responsive to player actions, leading to different outcomes and providing a more personalized experience. 

Given a sequence of events ($S$), the following metrics measure the difference between the subsequent events ($S'$) in response to players taking different actions after the leading sequence $S$. 

\begin{itemize}
\item {\bf Subsequent Plot Divergence}\hspace{2mm} We execute a pair of contrasting player actions after $S$, and compare how the following plot progression diverges semantically based on the player's different actions. The contrasting actions are attacking/killing a character (negative action) and seeking help for a character in danger (positive action). This comparison is performed by calculating two types of ROUGE distances. Specifically, the distance is computed pairwisely between the two plots generated after the positive and negative actions. A higher ROUGE distance indicates a greater divergence between the two plots driven by contrasting player actions, thus reflecting their higher impact on the plot progression. 
\item {\bf Perceived World State Change}\hspace{2mm} The metric assesses the perceived alternation of world states caused by the player's actionn in a specific scenario. We execute a player action of killing a character, and count the frequency of the killed character's reappearance in the subsequent plots.
\item {\bf Player Character Involvement}\hspace{2mm} The metric examines whether the player's action increases the player character's involvement in the subsequent events in the plot. We calculate the frequency with which the player's character appears in the plots that follow the positive action of helping a character. This indicates the extent to which the player's actions influence their engagement in the narrative.
\end{itemize}




We use the \presentation{\textit{``Fairytale Forest''} story domain (Figure~\ref{baseline})} and a fixed story outline containing two acts for evaluation. The player character is set as the dove. To initialize, we generate the plot for the first act in the outline as $S$ and set the world state accordingly. We then execute the designed player actions, and then use our method and the baseline method to generate a batch of 20 plots following each player action to compute the above metrics.

As Table~\ref{tech_eval:action_impact} shows, our approach generates significantly more diverse plots following the player's contrasting actions. Moreover, our approach generates plot with better perceived world state change and character involvement following player's actions. Overall, \presentation{we found that \textsc{WhatELSE} integrates player actions with a higher impact in the narrative generation process.}




%% file: sections/discussion.tex
The rise of generative models and prompt engineering has significantly impacted various fields and domains. Our user study results suggest that even laypeople can effectively use LLMs to create executable game plots for interactive narratives with appropriate interactive support. Therefore, it is valuable to continue exploring new interaction designs that assist novice users in creating interactive narratives. In this section, we discuss the implications derived from our system design and user studies, as well as the limitations and potential directions for future work.

\subsection{Revisiting the Design Goals}
We revisit our design goals presented in Section~\ref{design_goals} and reflect on the degree to which they have been achieved. We also discuss the opportunities and challenges for designing future AI-bridged IN authoring systems.

\textbf{DG1: Enable users to perceive the narrative space} {\sc WhatELSE} provides three views for the user to perceive the narrative space: pivot, outline, and variants view. The pivot view displays a user-defined narrative instance as the representative example in the narrative space. In the study, we observed most participants (9/12) did not change the pivot upon generating the outline. Instead, they alternated between variants and outline view to edit the narrative space. Particularly, they found variants view useful in providing {\it ``inspiration''} (P2) or surprising instances, indicating people may \presentation{underestimate the size of narrative space} when authoring IN. Participants also used variant views to prevent instances from deviating from their authorial intent. The variants view shows the distance of each variant to the pivot, allowing users like P1 to quickly identify outliers and remove unexpected instances in their narrative space. Currently, the variants view displays variants based on authorial intent distance and emergent behavior distance from the pivot. It would be interesting to let users customize these dimensions and dynamically visualize the narrative space, similar to Luminate \cite{suh2024luminate}.

\textbf{DG2: Support configurable level of abstraction in editing narrative space} \presentation{Our tools give users the flexibility to adjust the level of detail in their authoring of the narrative space. With the abstraction ladder, they can configure the overall outline, while the abstraction tooltip lets them fine-tune sentences or individual words.}
This granular control helps users create the narrative space more effectively. The user study reflected this with higher ratings on user control (Figure~\ref{fig:questionnaire}.a) and positive interview responses. Additionally, the technical evaluation results in Table~\ref{tech_eval:abstraction_ladder} demonstrate the abstraction ladder's effectiveness in managing different levels of abstraction. In the user study, participants with narrative writing experience easily understood the abstraction ladder levels, referring to a level as {\it ``story beats''} (P2). Other participants needed more exploration to find their desired level of abstraction. While the abstraction ladder is designed based on narrative structure, future work should explore making the levels more intuitive for people unfamiliar with narrative writing.

\textbf{DG3: Generate meaningful game events that react to player actions at play-time} We adopted an LLM-based narrative planning method that generates causally sound character actions that act out the event defined in the outline. In the study, participants were more satisfied with the game events generated by {\sc WhatELSE} compared to the baseline (Figure~\ref{fig:questionnaire}.c). They found that the choices they made in the game had a more realistic impact on the characters. Furthermore, the results of the technical evaluation in Table~\ref{tech_eval:action_impact} demonstrate that {\sc WhatELSE} can generate diverse plots that respond to the player’s contrasting actions. Like other LLM-based narrative generation, the quality of generated plots is limited by the challenge of preserving long-term dependency and coherence \cite{mirowski2023co}. Future work should explore approaches such as increasing the LLM’s context window \cite{kaddour2023challenges} and adopting a hierarchical generation approach \cite{mirowski2023co}.


\subsection{Intuitive and Analytic Thinking in Authoring Narrative Space} 

\presentation{WhatELSE supports both intuitive and analytical thinking when authoring IN. 
Authoring IN with instances aligns with people's natural writing process. } It is also straightforward to judge the moral expression in a narrative instance. When authors begin with uploading a draft story, their edits and creations on narrative instances reflect intuitive storytelling, where specific events guide their thinking about what could happen next.  

More analytical thinking occurs when authors abstract the outline to shape the narrative space. The act of replacing specific characters or events with abstract descriptions, such as changing {\it "the hunter"} to {\it "a human character with power"}, \presentation{is a deliberate attempt}. This process of using language abstraction to incorporate greater diversity requires \presentation{considerations and planning}. By gradually refining the outline through abstraction, authors weigh different narrative potentials and ensure that the result still conveys the desired moral.

{\sc WhatELSE} enables creators to jointly leverage both modes of thinking in their creation. By starting with concrete narrative instances with clear perception, creators can quickly ground their ideas and establish the foundation for the narrative space to shape. Through the process of abstraction, they deliberately decide operations to shape the narrative space. This transition between intuitive and analytical thinking may serve as a plausible explanation for {\sc WhatELSE}'s advantage for authors to maintain control over the narrative space while allowing for creative flexibility.

\subsection{Mixed-initiative Design for LLM Applications}

Participants' feedback highlights that, compared to fully free-form conversational assistants, \textsc{WhatELSE} is more helpful for novices in creating their games. As P12 noted, {\it ``I can explore what the different versions (abstraction levels) look like''}, and P6 added, {\it ``(WhatELSE) allows you to explore more.''} These comments point to a deeper reason behind the effectiveness of {\sc WhatELSE}. Though conversational-based chatbots seem to provide infinite options to choose from in creation, they are overwhelming to novices. P1 has been acutely aware of this challenge and left an opinion: {\it ``Maybe someone who is an experienced writer could be like.. I want to structure the plot this way. But as me, a person who isn't a writer, it's hard to really know what to do (with the baseline)''}. 

In contrast, the mixed-initiative design of \textsc{WhatELSE} primes users to explore more in sculpting the narrative space in their iterative trials by implicitly guiding them in a structured workflow of outline generation, of which P14 named as {\it ``indicators of what I'm trying to do''}. P2, who has moderate experience in narrative writing, also noticed the design and said, {\it ``It's very close to how I think when I write a story, those levels, those metrics''}.

\subsection{Strategic Integration of \textsc{WhatELSE} in Interactive Narrative Creation}

A number of IN authoring tools have been developed, including branching-based systems like Twine \cite{friedhoff2013untangling} and event-node graph frameworks \cite{chung2024patchview}. \textsc{WhatELSE} is designed to complement, rather than compete with, these existing tools. We aim to enable \textsc{WhatELSE} to augment existing tools by addressing the challenges widely applied to interactive narrative creation so that a more comprehensive workflow can be supported.

\presentation{\textsc{WhatELSE} can be used in combination with traditional authoring tools in several ways.}
For example, the variants generated by \textsc{WhatELSE} can be converted into event diagrams or storyline branches. Therefore, the shaped narrative space via the narrative space editor can be exported to traditional interactive narrative authoring tools for fine-grained manual editing. Similarly, the fully structured representations of the narrative space can be exported to the interactive narrative compiler, as a clearer guide used in the plot execution mode.

Furthermore, our system aims to address the broader challenges of AI-bridged creation of interactive narratives. Within this paradigm, \textsc{WhatELSE} provides creators with the tools to shape, refine, and manage the narrative space. This concept is closely connected to the design space explored in other creative domains; thus, \presentation{studies that investigate the design exploration and management} in design space can be further adapted to this specific application for narrative space editing~\cite{ez2022design,suh2024luminate}, with their proposed techniques providing another view of the narrative space.



\subsection{Branching Outline with Multiple Pivots}

\begin{figure*}
    \centering
    \includegraphics[width=.98\textwidth]{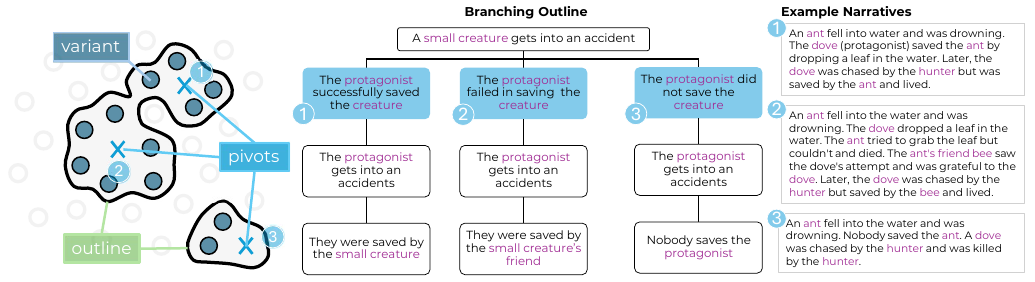}
    \caption{\presentation{Illustration of a narrative space generated by a branching outline with three pivots to indicate three types of storylines.}}
    \label{fig:multipivot}
    \Description{}
\end{figure*}

It is possible to combine the traditional branching storyline graph representation with our outline representation, to obtain branching outline graphs where each node represents an abstract event. Figure \ref{fig:multipivot} shows such a branching outline, where each path in this graph corresponds to multiple concrete storylines.

One potential extension of this work is to generalize our workflow and system to support creating such branching outline. The more general system should allow the user to provide multiple narrative examples as input to initialize the narrative space, and multiple instances in the narrative space should be able to serve as pivots. 
\usecase{For instance, the user can provide three example narratives as pivots to indicate three types of storylines \presentation{(Figure \ref{fig:multipivot})}}.


Branching outline representation expands the expressivity of both our outline representation and traditional branching graph representation. Our system has shown a linear outline can be unfolded to instantiate the same story structure in multiple concrete forms. A branching outline representation can potentially support organizing multiple story structures together for expressing more complex themes. However, it also raises new challenges for both technical pipeline and interface design, for example, how to computationally construct branching outlines from multiple example narratives, and how to assist the user in better perceiving connection between different narrative instances from the same and different abstract branches, especially when the boundary between different abstract branches can be blurred due to the fuzziness of the trigger conditions expressed with abstract language.

\section{Limitation and Future Work}
\label{sec:limitation}

We describe the limitations of our work to clearly define the scope of our findings and inspire future research directions.


\paragraph{Study Design} We evaluated our system through an in-person study \presentation{with 12 participants who have minimal IN authoring experience and moderate generative AI experience. The study task used a simple story domain with a template story to minimize learning burdens}. In real-world scenarios, IN creation typically involves a larger narrative space with more complex plots. Therefore, studies that use more complex creation tasks and have a more diverse participant background, such as including people with limited AI experience or professional IN writers, would give us a deeper understanding of the usability, effectiveness, and generalizability of our work. \evaluation{Additionally, {\sc WhatELSE} has not been directly compared with traditional IN authoring tools. While a full comparison of {\sc WhatELSE} and several traditional tools across an entire IN authoring process may not be an overkill, experiments targeting specific perspectives, such as branching capability, can better clarify the unique contribution or limitation of {\sc WhatELSE}. }

\evaluation{\paragraph{Player Experience} Although this work focuses primarily on the author's experience, the player experience is also important. While the results of the technical evaluation show that {\sc WhatELSE}  generates diverse plots that respond to simulated player’s actions, future studies the collect subjective player feedback, such as their engagement and enjoyment through crowdsourced assessment of text-adventure games generated using {\sc WhatELSE}, would help us better understand this system.  
}

\paragraph{Interface Design} 
The system provides the pivot, variant, and outline view to help perceive the narrative space. Within these three views, more intuitive feedback could be introduced such as providing quantifiable measures of abstraction and concreteness in the outline, similar to the \presentation{metrics of authorial intent and emergence for variants.} While the outline view depicts the boundary of the narrative space, the outline itself is a free-form natural language query. By incorporating classic views of IN, such as a branching diagram, we could explore using a semi-structured outline that combines the flexibility of natural language with the clarity of structured representations. Additionally, introducing an intermediary component between the mutual transformation between narrative instances and the outline, such as an event diagram, would create a "trinity" of instance, semi-structure, and outline, offering more control and granularity in narrative creation.

\paragraph{Export the Game Plot} One of the key advantages of our approach is that game plot is generated through narrative planning, making it fully structured and controllable. \presentation{It is possible to deploy the generated game plot into a real game engine so that each event in the plot corresponds to an in-game function call.} 
Future work will leverage this advantage to use an existing game engine to serve as an intermediate between \textsc{WhatELSE} and game players, where both the game plots and player actions will operate the game engine, embodying the generated game plot in a real gaming experience.

\section{Conclusion}

Generative AI advances interactive narrative creation by enabling just-in-time content generation that adapts to player choices. However, this increased interactivity makes it difficult for authors to control the narrative space. In this paper, we introduced WhatELSE, an interactive narrative authoring system that tackles this challenge via a mutual transformation between narrative instances and narrative space. Through its three views—narrative instance, outline, and variants—WhatELSE empowers authors to perceive and shape narrative boundaries using linguistic abstraction. By leveraging an LLM-based simulation, \textsc{WhatELSE} further unfolds narrative spaces into executable gameplots. Our user study (N=12) and technical evaluations showed that \textsc{WhatELSE} enables the creation of structured yet flexible gameplots, making it an effective tool for interactive narrative creation. We believe our work contributes to advancing creators' balance of their authorial intent and player interactivity in AI-bridged interactive narrative creation.